\documentclass[twocolumn,aps,prl,floatfix,longbibliography,groupedaddress]{revtex4-1}
\usepackage{amsmath,bbm}
\usepackage{graphicx}
\usepackage{multirow}
\usepackage{hyperref}
\usepackage{amsfonts}
\usepackage{amsbsy}
\usepackage{xcolor}
\usepackage{soul}
\usepackage{graphicx}
\usepackage{mathtools}

\renewcommand{\vec}[1]{\mathbf{#1}}

\newcommand{\e}{\mathcal{E}}
\newcommand{\Pc}{\mathcal{P}}
\newcommand{\cbE}{\boldsymbol{\mathbf{\cal E}}}
\newcommand{\bra}[1]{\ensuremath{\left\langle #1 \right\vert}}
\newcommand{\ket}[1]{\ensuremath{\left\vert #1 \right\rangle}}

\renewcommand{\vec}[1]{\mathbf{#1}}

\begin{document}
\title{Optical Magnetism and Huygens' Surfaces in Arrays of Atoms Induced by Cooperative Responses }

\date{\today}
\author{K.~E.~Ballantine}
\email{k.ballantine@lancaster.ac.uk}
\author{J.~Ruostekoski}

\email{j.ruostekoski@lancaster.ac.uk}

\affiliation{Department of Physics, Lancaster University, Lancaster, LA1 4YB, United Kingdom}

\begin{abstract}
By utilizing strong optical resonant interactions in arrays of atoms with electric dipole transitions, we show how to synthesize collective optical responses that  correspond to those formed by arrays of magnetic dipoles and other multipoles. 
Optically active magnetism with the strength comparable with that of electric dipole transitions is achieved in collective excitation eigenmodes of the array.
By controlling the atomic level shifts, an array of spectrally overlapping, crossed electric and magnetic dipoles can be excited, providing a physical realization of a nearly-reflectionless quantum Huygens' surface with the full $2\pi$ phase control of the transmitted light that allows for 
extreme wavefront engineering even at a single photon level.
We illustrate this by creating a superposition of two different orbital angular momentum states of light from an ordinary input state that has no orbital angular momentum.

\end{abstract}

\maketitle

A crucial limitation for utilizing atoms as optical media is the inability of light to couple to atoms using both its electric and magnetic components. Optical magnetic dipole transitions in atoms typically are very weak to the extent that magnetic susceptibility at optical frequencies has generally been considered a meaningless concept~\cite{LandauED}, and the optical response is determined by strong oscillating electric dipoles~\cite{Cowan81}. The absence of magnetic coupling in natural media has led to the development of artificial metamaterials~\cite{Zheludev12} and metasurfaces~\cite{capasso_review}, representing man-made designer materials that simultaneously provide strong interactions with both field components of light~\cite{Novotny12}. 
The quest for dramatic consequences of this from perfect lensing~\cite{PendryPRL2000} to invisibility cloaks~\cite{PendryEtAlSCI2006,LeonhardtSCI2006,SchurigSci2006} has been driving the development, but performance at optical frequencies has been limited.

Here we show that strong light-mediated interactions between cold atoms in planar arrays can be designed to synthesize \emph{collective} radiative excitations that exhibit strong electric and magnetic optical responses. The system has considerable advances over artificial fabricated materials because of the absence of dissipative losses due to absorption and the possibility to reach quantum regime in the optical manipulation and control. We illustrate the flexibility of designing collective radiative excitations by proposing a Huygens' surface of atoms, where the crossed electric and magnetic dipolar resonances act as elementary sources for wave propagation according to the Huygens' principle~\cite{Huygens,Love1901}. 
The Huygens' principle then states that an \emph{arbitrary} wavefront can be constructed by an ideal physical realization of a Huygens' surface, therefore achieving extreme optical manipulation.
In a single-photon limit, our proposed array provides a quantum-photonic wave-engineering tool, with beam shaping, steering, and focusing capabilities. 

Atomic physics technology provides a variety of approaches for trapping closely-spaced atoms in arrays with single-site control and unit-occupancy per site~\cite{Weitenberg11,Lester15,Xia15,Endres16,Barredo16,Kim16,Cooper18}. For cold atoms with subwavelength spacing, the light-mediated interactions can be very strong due to resonant light undergoing multiple scattering between the atoms~\cite{Jenkins2012a,Perczel2017a,Bettles2017,Bettles2016,Facchinetti16,Facchinetti18,Shahmoon,
Plankensteiner2017,Asenjo-Garcia2017a,Jen17,Guimond2019,Ritsch_subr,Kramer2016,Sutherland1D,Yoo2016,Zhang2018,Mkhitaryan18,Bhatti18,Henriet2018,Plankensteiner19,Javanainen19,Bettles20,Qu19,Zhang20}, when the realization of strong collective coupling in large resonator arrays of metasurfaces typically requires special arrangements~\cite{lemoult2010,Jenkins17}. Moreover, the atomic arrays can offer a promising platform for quantum information processing at the level of single photon excitations~\cite{Grankin18,Guimond2019,Ballantine20ant}.
Experiments on strong collective optical responses of trapped cold atomic ensembles are actively ongoing~\cite{BalikEtAl2013,CHA14,Pellegrino2014a,Havey_jmo14,wilkowski,
Jennewein_trans,Ye2016,Jenkins_thermshift,vdStraten16,Guerin_subr16,Machluf2018,Dalibard_slab,Bettles18}, and the first measurements of the transmitted light through an optical lattice of atoms in a Mott-insulator state have now been performed~\cite{rui2020}
that demonstrate subradiant resonance narrowing where the entire lattice responds as a coherent collective entity.  

Here we propose to engineer strong resonant dipole-dipole interactions by design atomic arrays, such that each unit-cell of the periodic lattice, consisting of a small number of sites, has a specific symmetry that characterizes the collective excitation eigenmodes of the array. The modes extend over the entire sample and in large lattices the effect of the light-mediated interactions becomes strong, even when radiation rates  of an individual unit-cell are closer to those of an isolated atom. This allows us to utilize strongly subradiant collective eigenmodes, with suppressed emission rates, as weakly radiating `dark' states.  We use the dark states to synthesize optically active magnetism via an array of magnetic dipoles that can indirectly be excited by incident light.

We show that collective resonances can be superposed to form an array of electric and magnetic dipoles. Together with the control of the full $2\pi$ phase coverage of transmitted light with almost no reflection,
this realizes a Huygens' surface for light. Wavefront-engineering is illustrated by transforming a Gaussian wave to a superposition of orbital angular momentum (OAM) beams with differing OAM. 
In the single-photon limit, such entangled OAM states can provide a useful resource for quantum information~\cite{Mair01} that is not limited to two-state systems~\cite{Bechmann00}.
Physical implementations of Huygens' surfaces in weakly interacting metamaterials have been achieved in metallic systems using microwaves~\cite{Pfeiffer13}, but optical frequencies pose severe challenges, e.g., due to absorption and lack of suitable magnetic
resonances. Near-infrared/optical Huygens' sources or closely-related metasurface controls have been realized using dielectric nanoparticles~\cite{Decker15,Yu15,Chong15,Shalaev15,Decker15,Kruk16,Lukyanchuk16}.

\begin{figure}[htbp]
  \centering
   \includegraphics[width=\columnwidth]{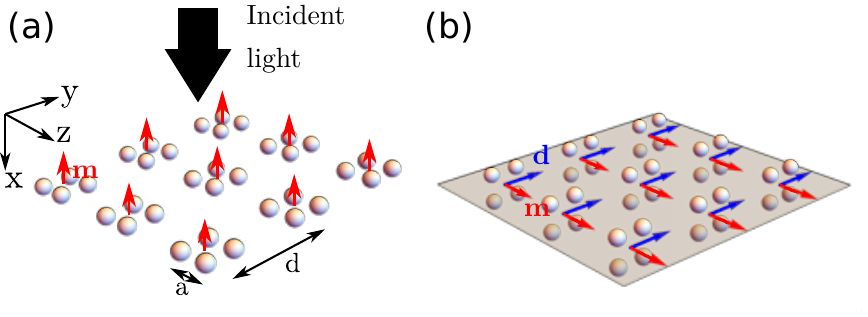}
  \caption{(a) Array with square unit-cells of lattice constant $d$ and unit-cell of size $a$ to generate an array of magnetic dipoles pointing normal to the plane. (b) Bilayer array forming a Huygens' surface with crossed electric ($\vec{d}$) and magnetic ($\vec{m}$) dipoles.   }
  \label{fig1}
\end{figure}

The atoms are confined in a 2D lattice in the $yz$ plane with one atom per site. We consider a $J=0\rightarrow J^\prime=1$ transition with a controllable Zeeman splitting of the $J^\prime=1$ levels, which could be achieved with ac Stark shifts of lasers or microwaves~\cite{gerbier_pra_2006} or magnetic fields. The formalism we employ can describe the dynamics of two different regimes. The first is the decay of a single-photon excitation spread across the lattice, described by the density matrix $\rho=\ket{\psi}\bra{\psi}+p_g\ket{G}\bra{G}$ where $p_g$ is the probability that the excitation has decayed, $\ket{G}$ is the state with all atoms in the ground state and single-excitation states are represented by 
$
\ket{\psi}= \sum_{j,\nu} \Pc^{(j)}_{\nu}(t)\,\hat{\sigma}^{+}_{j\nu}\ket{G}
$~\cite{SOM}.
Here $\hat{\sigma}_{j\nu}^{+}$ is the raising operator from the ground state to the excited state $\nu$ on atom $j$~\cite{SVI10,Needham19,Grankin18,Ballantine20ant}. Such a single-photon excitation could be initialized by short-range coupling to a control qubit, e.g., using laser-assisted coupling between Rydberg states~\cite{Grankin18}.

The same dynamics (when only considering one-body expectation values) also describes the classical polarization amplitudes of each atomic dipole in the limit of low drive intensity~\cite{Morice1995a,Ruostekoski1997a,Javanainen1999a,Sokolov2011,Javanainen2014a,Sutherland16forward}. The dipole moment of atom $j$ for the transitions $\ket{J=0,m=0}\rightarrow\ket{J^\prime=1,m=\sigma}$ is $\vec{d}_j=\mathcal{D}\sum_{\sigma}\hat{\vec{e}}_\sigma\mathcal{P}_\sigma^{(j)}$, where $\mathcal{D}$ denotes the reduced dipole matrix element, $\mathcal{P}_\sigma^{(j)}$ the polarization amplitudes, and $\hat{\vec{e}}_\sigma$ unit polarization vectors~\cite{SOM}. In this case, the atoms are typically driven by a laser.
We take an incident light field as $\cbE(r)=\e_0(y,z)\hat{\vec{e}}_y \exp{(ikx)}$ \footnote{The light and atomic field amplitudes here refer to the slowly varying positive frequency components, where the rapid oscillations $\exp(-i \omega t)$ at the laser frequency have been factored out.}.

In a vector form $\vec{b}_{3j-1+\sigma}=\mathcal{P}_\sigma^{(j)}$, the collective response of atoms then results from a linear set of coupled equations due to the driving by the incident light, $\vec{F}_{3j-1+\sigma}=i\xi\hat{\vec{e}}_\sigma^\ast\cdot\epsilon_0\cbE(\vec{r}_j)/\mathcal{D}$ where  $\xi=6\pi\gamma/k^3$ is given in terms of  the single atom linewidth $\gamma=\mathcal{D}^2k^3/(6\pi\hbar\epsilon_0)$, and the scattered light from all the other atoms, such that
$
\dot{\vec{b}} = i\mathcal{H}\vec{b}+\vec{F}
$~\cite{Lee16,SOM}.
The matrix $\mathcal{H}$ has diagonal elements $\Delta_\sigma^{(j)}+i\gamma$, where $\Delta_\sigma^{(j)}$ is the detuning of level $m=\sigma$ from resonance with the driving frequency $\omega$. The off-diagonal elements correspond to dipole-dipole coupling between atoms $\xi\hat{\vec{e}}_\sigma^\ast\cdot\mathsf{G}(\vec{r}_j-\vec{r}_l)\hat{\vec{e}}_{\sigma^\prime}$, for $j\neq l$,  where $\mathsf{G}$ denotes the dipole radiation kernel, such that $
\epsilon_0 \vec{E}_s^{(l)}(\vec{r})=\mathsf{G}(\vec{r}-\vec{r}_l)\vec{d}_l
$
gives the scattered field at $\vec{r}$ from the atom $l$ at $\vec{r}_l$~\cite{Jackson,SOM}.
The response can then be understood in terms of the collective eigenmodes $\vec{v}_n$ of non-Hermitian $\mathcal{H}$~\cite{Jenkins_long16}, with corresponding eigenvalues $\delta_n+i\upsilon_n$, where $\delta_n=\omega_0-\omega_n$ and $\upsilon_n$ denote the collective line shift and linewidth, and $\omega_0$ is the unshifted transition resonance frequency.

\begin{table}\caption{\label{squaremodes} Normalized multipole decomposition of the eigenmodes of a square unit-cell with $a=0.15\lambda$ and $\Delta_\sigma^{(j)}=0$, as well as collective linewidths $\upsilon_{i}^{\mathrm{uc}}$ of each mode in an isolated unit-cell and $\upsilon_{i}$ in a $20\times 20$ lattice of unit-cells with $d=0.5\lambda$. In (1-6) the polarization is in the $y z$ plane while in (7-9) along the $x$ direction. Degenerate modes are labeled with their degeneracy.}
\begin{ruledtabular}
\begin{tabular}{lcccccc}
 \hline
   $n$ & $|\vec{d}|$ & $|\vec{m}|$ & $|\vec{q}_\mathrm{e}|$ & $|\vec{q}_\mathrm{m}|$ & $\upsilon_{i}^{\mathrm{uc}}/\gamma$ & $\upsilon_{i}/\gamma$ \\ \hline

 1 (x2) & 1 & & & & 3.4 & 3.7 \\ 
 2 & & 1 & & & 0.4  & $10^{-4}$ \\ 
 3 & & & 1 & & 0.25  & $10^{-3}$ \\ 
 4 & & & 1 & & 0.25 & $10^{-3}$ \\ 
 5 & & & 1 & & 0.08 & $10^{-3}$ \\ 
 6 (x2) & 0.89 & & & 0.05 & 0.09 & 0.1 \\ \hline
 7 & 1 & & & & 3.3 & 0.02 \\ 
 8 (x2) & & 0.63 & 0.37 & & 0.3 & $10^{-4}$ \\ 
 9 & & & & 0.78 & 0.02 & $10^{-8}$ \\ 
 \hline
\end{tabular}
\end{ruledtabular}
\end{table}

We engineer the spatially extended collective eigenmodes of a large array by designing the light-mediated interactions between the atoms in terms of the symmetries of an individual unit-cell at each site that forms the lattice. Then the different eigenmodes of an appropriately chosen unit-cell correspond to different multipole excitations, such as electric dipole, magnetic dipole, electric quadrupole, etc. 
Moreover, radiative interactions between atoms lead to collective eigenmodes of the entire array which behave as an effective lattice of unit-cells, each with radiative properties determined by  these multipole moments. Such eigenmodes can be excited by selectively manipulating each unit-cell.

We consider first a square array of unit-cells in the $yz$ plane with lattice spacing $d$, where each unit-cell consists of four atomic sites forming a square with side-length $a$, as illustrated in Fig.~\ref{fig1}(a).  
To characterize the optical response, we take the far-field radiation from each unit-cell and decompose it into vector spherical harmonics~\cite{SOM}
\begin{equation}
\vec{E}_s^{(j)} = \sum_{l=0}^\infty\sum_{m=-l}^{l}\left(\alpha_{\mathrm{E},lm}^{(j)}\vec{\Psi}_{lm}+\alpha_{\mathrm{B},lm}^{(j)}\vec{\Phi}_{lm}\right).
\label{eq:harmonics} 
\end{equation}
This expansion allows a collection of point dipoles with different positions and orientations to be decomposed into a set of multipole moments at the origin~\cite{Jackson}, 
with $\vec{\Psi}_{lm}$ corresponding to an electric dipole $\vec{d}$ for $l=1$, quadrupole $\vec{q}_{\mathrm{e}}$ for $l=2$, octopole for $l=3$, etc., while $\vec{\Phi}_{lm}$ represents magnetic multipoles (dipole $\vec{m}$, quadrupole $\vec{q}_{\mathrm{m}}$, etc.).

\begin{figure}[htbp]
  \centering
   \includegraphics[width=\columnwidth]{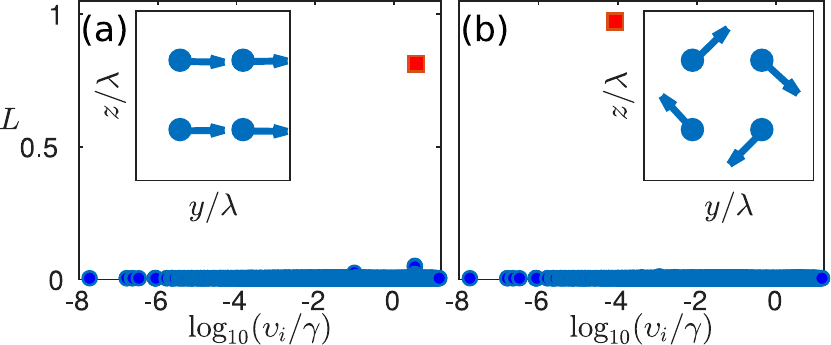}
  \caption{(a) Occupations of collective excitation eigenmodes (ordered by linewidth $\upsilon_i$) in a steady state for a 20$\times$20($\times$4) array ($d=0.5\lambda$, $a=0.15\lambda$), shown in Fig.~\ref{fig1}(a), illuminated by a Gaussian profile $\e_0(y,z)$ with $1/e^2$ radius $7\lambda$, when detunings are optimized to target the collective eigenmode (red square) which is closest to a uniform repetition of the unit-cell electric dipole mode.  (b) As in (a), but with a target magnetic dipole eigenmode closest to a uniform distribution of magnetic dipoles in the $x$ direction (red square).
Insets (a,b): the resulting atomic dipoles on a central unit-cell.   }
  \label{figeb}
\end{figure}

To analyze the light-mediated interactions in a single unit-cell which makes up the lattice, we take the eigenvectors $\vec{v}_n$ of the matrix $\mathcal{H}$ corresponding to these four atoms in isolation, and decompose the scattered field for each of these eigenmodes. The results, normalized so that the sum of all multipole moments is one for each mode, are listed in Table~\ref{squaremodes} up to quadrupole moments, with the corresponding eigenmodes illustrated in Fig.~S1~\cite{SOM}. Most modes are dominantly represented by one of the multipole components and we find, e.g., eigenmodes with electric dipoles in the $y$ or $z$ (n=1, doubly degenerate), or $x$ (n=7) directions. More interestingly, however, there are collective eigenmodes almost solely corresponding to a magnetic dipole in the $x$ direction (n=2) [inset of Fig.~\ref{figeb}(b)], electric quadrupoles, and those dominated by magnetic quadrupoles.

The resonances of electric dipole eigenmode (EDM) and magnetic dipole eigenmode (MDM) are only shifted by $3.8\gamma$---on the order of the collective linewidth $3.4\gamma$ of the superradiant EDM (Table~\ref{squaremodes}). The MDM is subradiant (linewidth $0.4\gamma<\gamma$). For each of these modes, there is a corresponding collective mode of the whole lattice which best resembles a uniform repetition of the eigenmode of each unit-cell. The collective interacting nature of the eigenmodes in large arrays is most clearly manifested in the strongly subradiant response, with several of the modes having linewidths orders of magnitude narrower than that of a single atom, or an isolated unit-cell.

The collective modes can be excited by varying the atomic level shifts within each unit-cell, as illustrated in Fig.~\ref{figeb}. We optimize the level shifts on each atom in the unit-cell to maximize overlap with the desired mode, while keeping each unit-cell identical. Targeting the EDM of Fig.~\ref{figeb}(a) is straightforward  with the incident field propagating along the $x$ axis and linearly polarized along the $y$ axis.  The occupation measure of the collective eigenmode $\vec{v}_j$ is defined as $L_j = {|\vec{v}_j^T \vec{b}(t)|^2/\sum_i|\vec{v}_i^T \vec{b}(t)|^2}$~\cite{Facchinetti16}, with the resulting dominant occupation of the array-wide EDM closest to a uniform repetition of the unit-cell EDM. The central unit-cell clearly shows the collective electric dipole moment [inset of Fig.~\ref{figeb}(a)].

While the uniform collective EDM of the whole array is phase-matched with the incident light  
and is easy to excite, there is no direct coupling of the incident field to the uniform MDM. The MDM is also subradiant (`dark') and difficult to excite. 
By choosing appropriate level shifts $\Delta_\sigma^{(j)}$ for individual atoms, we can break the symmetry, such that the EDM and MDM are no longer eigenmodes of the collective light-atom system of vanishing
level shifts. The level shifts induce a coupling between the EDM and MDM, therefore exciting the MDM by first driving the EDM by the incident field, followed by the transfer of the excitation to the MDM. The process can be illustrated by an effective two-mode model between the two collective modes~\cite{SOM}
\begin{subequations}
\begin{align}
\label{zeeman1a}
\partial_t \mathcal{P}_{e} &= (i\delta_e+i\Delta-\upsilon_e)\mathcal{P}_{e} + \delta\mathcal{P}_{m} +f, \\
\label{zeeman2a}
\partial_t \mathcal{P}_{m} &= (i\delta_m+i\Delta-\upsilon_m)\mathcal{P}_{m}+ \delta\mathcal{P}_{e},
\end{align}
\label{zeeman}
\end{subequations}
where  $\mathcal{P}_{e,m} $ are the amplitudes, and $\delta_{e,m}$ and $\upsilon_{e,m}$ the collective resonance line shifts and linewidths of the EDM and MDM, respectively, and  $f$ denotes the incident-field driving that is only coupled to $\mathcal{P}_{e} $.
The contribution of the alternating level shifts in a unit-cell is encapsulated in $\delta$~\cite{SOM} which induces a coupling between the modes. The alternating level shifts can be generated, e.g., by the ac Stark shifts of crossed standing waves~\cite{SOM}.

The effective dynamics of Eq.~\eqref{zeeman} can represent both a single unit-cell in isolation and the entire array of multiple unit-cells. The two cases dramatically differ by the value of $\upsilon_m$ that  becomes strongly subradiant as the size of the array increases ($\upsilon_m\simeq 10^{-4}\gamma$ for a 20$\times$20 array, Table~\ref{squaremodes}; see also~\cite{SOM}). The ratio between the occupations of the MDM and EDM in the steady state of Eq.~\eqref{zeeman} at the resonance of the MDM is $\delta^2/\upsilon_m^2$. Therefore, the narrow $\upsilon_m$ in large arrays  allows the MDM to dramatically dominate the EDM, even for relatively small level splittings $\delta$, providing a protocol for synthesizing a magnetic dipolar response that utilizes strong cooperative interactions in large systems. In other words, because of the deeply subradiant nature of the MDM, this excitation does not decay and becomes dominant in the steady state, in a process reminiscent of the electromagnetically-induced transparency~\cite{FleischhauerEtAlRMP2005} of `dark' and `bright' states of noninteracting atoms.
The excitation of the uniform collective MDM is shown in Fig.~\ref{figeb}(b) with an occupation of $\approx 0.97$. For a single unit-cell, the amplitude of the magnetic dipole radiation here is about 23\% of that of a single atom.

\begin{figure}[htbp]
  \centering

   \includegraphics[width=\columnwidth]{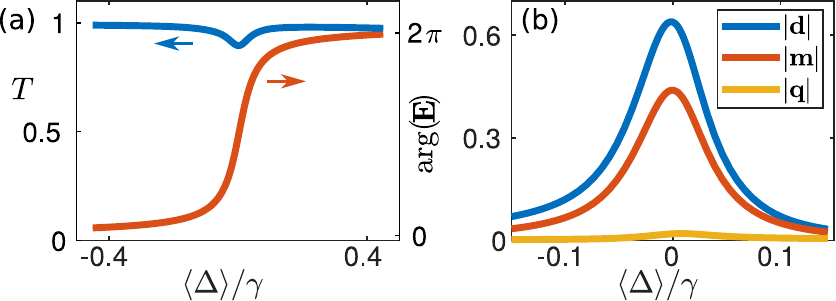}
  \caption{(a) Transmission $T$ (left axis) and phase (right axis) of the incident and scattered light in the forward direction as a function of the laser frequency for plane-wave illumination of a $32\times 32$ lattice ($d=0.8\lambda$, $a=0.15\lambda$) showing the full $2\pi$ range with transmission everywhere exceeding $90\%$. (b) Decomposition of the corresponding radiative excitation into (curves from top to bottom) electric and magnetic dipole, and electric quadrupole contributions. The relative level shifts of all levels are held fixed, and identical on each unit-cell. 
  }
  \label{fighlattice}
\end{figure}

We next utilize the combination of electric and magnetic dipoles to prepare a Huygens' surface. 
Huygens' principle states each point on a propagating wave acts as a source of secondary spherical waves which interfere to produce the subsequent wavefront. A more rigorous formulation models each point as 
a perpendicular pair of electric and magnetic dipoles of equal strength~\cite{Love1901,Schelkunoff36}. Each such point then acts as an ideal point source for light which propagates only in the forward direction. While these secondary sources were originally introduced as a conceptual means to understand wave propagation, a physical implementation of such a surface consisting of real emitters can be used to engineer arbitrary wavefronts by controlling the phase of the light produced at each point.

To form an effective Huygens' surface of crossed electric and magnetic dipoles  we consider a geometry where each square unit-cell is rotated around the $y$ axis and lies in the $xy$ plane, with the electric (magnetic) dipole in the $y$ ($z$) direction [Fig.~\ref{fig1}(b)]. This is equivalent to having a bilayer array with two sites per unit-cell in each layer. 
We can again identify collective eigenmodes that are characterized by a uniform array of either electric and magnetic dipoles. In this case the collective linewidth of the MDM remains broader as the size of the array increases, since all the dipoles are in the same plane.
We then control the atomic level shifts to form a superposition of the EDM and MDM on each unit-cell.

\begin{figure}[htbp]
  \centering
   \includegraphics[width=\columnwidth]{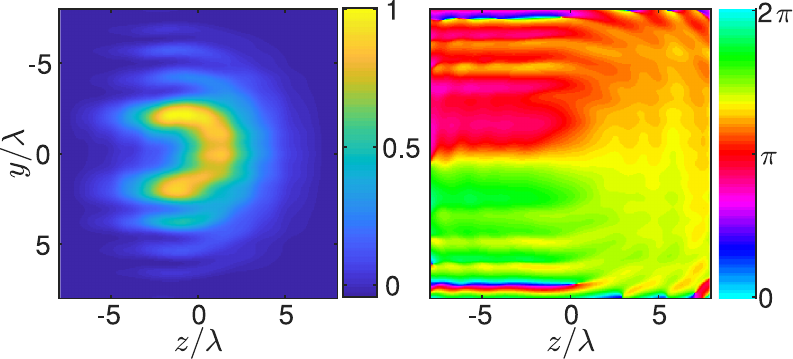}
  \caption{ (a) Transmitted intensity and (b) phase of the total electric field, calculated from each dipole at a distance $5\lambda$ for a $20\times 20$ lattice ($d=0.8\lambda$, $a=0.15\lambda$) shown in Fig.~\ref{fig1}(b). An incident Gaussian beam with waist $w_0=5\lambda$ is transformed into an equal superposition of states with orbital angular momentum $0$ and $\hbar$ per photon.  The intensity varies with angle due to interference between the two components. }
  \label{figoam}
\end{figure}

To utilize the Huygens' surface for wavefront engineering, we must control the phase of the total transmitted light at each unit-cell, which consists of both the incident and scattered light~\cite{SOM}. We show the control of the phase over the whole range of $0\leq{\rm arg}(\vec{E}) \leq 2\pi$ as a function of the detuning in an array of identical unit-cells  in Fig.~\ref{fighlattice}(a), while the transmission remains above $90\%$ at all points. Minimum transmission generally increases with the number of atoms, but is already over $80\%$ for a $20\times 20$ array. 
Note that spectrally nonoverlapping resonances of the EDM and MDM would produce significantly reduced transmission values at different frequencies. Having the two resonances spectrally overlapping yields a nearly flat line close to unity, despite providing a dramatic $2\pi$ phase change which is two times the phase shift that the magnetic or electric dipoles can individually contribute in resonance.
The multipole decomposition of a unit-cell at the center of the lattice is plotted in Fig~\ref{fighlattice}(b), showing the combination of electric and magnetic dipoles. The results in Fig.~\ref{fighlattice} are achieved by controlling the level shifts and also allowing the positions of each atom in the unit-cell to vary, up to a maximum of $0.05\lambda=a/3$, while keeping them constant from one unit-cell to another. 

We demonstrate wavefront-engineering with the Huygens' surface by transforming an ordinary input state with no OAM into an OAM state~\cite{SOM} or a superposition of OAM states, shown in Fig.~\ref{figoam} for 
 $(\ket{l=0}+\ket{l=1})/\sqrt{2}$, with the states of $l\hbar$ OAM per photon. In the single-photon limit, such entangled OAM states have applications to quantum information~\cite{Mair01}, and provide a larger alphabet $l=0,
 \dots,N$ for quantum information architectures~\cite{Bechmann00} than traditional two-state systems.
We take the relative level shifts to be equal on all unit-cells, while the overall shift varies such that the phase of the transmitted light gives the desired profile. 
 Only $2\%$ of the total power incident on the lattice is reflected backwards. 

In conclusion, we showed how to harness light-mediated resonant dipole-dipole interactions in atomic arrays to design collective excitations that exhibit optically active magnetism and a Huygens' surface.
This is achieved by first designing the interactions in an individual unit-cell of the array to engineer a lattice of magnetic dipoles, or of crossed spectrally-overlapping electric and magnetic dipoles.
The latter can be used to create a nearly-reflectionless Huygens' surface, consisting of ideal emitters with a fully $2\pi$-controllable phase, that allows for arbitrary shaping of wavefronts, including even those for a single photon.

 \begin{acknowledgments}
We acknowledge financial support from the UK EPSRC (Grant Nos.\ EP/S002952/1, EP/P026133/1). We have become aware of a related parallel theoretical proposal on generating magnetic dipoles for atoms in Ref.~\cite{Alaee20}.
\end{acknowledgments}


\begin{thebibliography}{91}%
\makeatletter
\providecommand \@ifxundefined [1]{%
 \@ifx{#1\undefined}
}%
\providecommand \@ifnum [1]{%
 \ifnum #1\expandafter \@firstoftwo
 \else \expandafter \@secondoftwo
 \fi
}%
\providecommand \@ifx [1]{%
 \ifx #1\expandafter \@firstoftwo
 \else \expandafter \@secondoftwo
 \fi
}%
\providecommand \natexlab [1]{#1}%
\providecommand \enquote  [1]{``#1''}%
\providecommand \bibnamefont  [1]{#1}%
\providecommand \bibfnamefont [1]{#1}%
\providecommand \citenamefont [1]{#1}%
\providecommand \href@noop [0]{\@secondoftwo}%
\providecommand \href [0]{\begingroup \@sanitize@url \@href}%
\providecommand \@href[1]{\@@startlink{#1}\@@href}%
\providecommand \@@href[1]{\endgroup#1\@@endlink}%
\providecommand \@sanitize@url [0]{\catcode `\\12\catcode `\$12\catcode
  `\&12\catcode `\#12\catcode `\^12\catcode `\_12\catcode `\%12\relax}%
\providecommand \@@startlink[1]{}%
\providecommand \@@endlink[0]{}%
\providecommand \url  [0]{\begingroup\@sanitize@url \@url }%
\providecommand \@url [1]{\endgroup\@href {#1}{\urlprefix }}%
\providecommand \urlprefix  [0]{URL }%
\providecommand \Eprint [0]{\href }%
\providecommand \doibase [0]{http://dx.doi.org/}%
\providecommand \selectlanguage [0]{\@gobble}%
\providecommand \bibinfo  [0]{\@secondoftwo}%
\providecommand \bibfield  [0]{\@secondoftwo}%
\providecommand \translation [1]{[#1]}%
\providecommand \BibitemOpen [0]{}%
\providecommand \bibitemStop [0]{}%
\providecommand \bibitemNoStop [0]{.\EOS\space}%
\providecommand \EOS [0]{\spacefactor3000\relax}%
\providecommand \BibitemShut  [1]{\csname bibitem#1\endcsname}%
\let\auto@bib@innerbib\@empty
\bibitem [{\citenamefont {Landau}\ \emph {et~al.}(1984)\citenamefont {Landau},
  \citenamefont {Lifshitz},\ and\ \citenamefont {Pitaevskii}}]{LandauED}%
  \BibitemOpen
  \bibfield  {author} {\bibinfo {author} {\bibfnamefont {L.~D.}\ \bibnamefont
  {Landau}}, \bibinfo {author} {\bibfnamefont {E.~M.}\ \bibnamefont
  {Lifshitz}}, \ and\ \bibinfo {author} {\bibfnamefont {L.~P.}\ \bibnamefont
  {Pitaevskii}},\ }\href@noop {} {\emph {\bibinfo {title} {Electrodynamics of
  Continuous Media, 2nd ed.}}}\ (\bibinfo  {publisher} {Pergamon Press, New
  York},\ \bibinfo {year} {1984})\BibitemShut {NoStop}%
\bibitem [{\citenamefont {Cowan}\ and\ \citenamefont
  {of~California~Press}(1981)}]{Cowan81}%
  \BibitemOpen
  \bibfield  {author} {\bibinfo {author} {\bibfnamefont {R.D.}\ \bibnamefont
  {Cowan}}\ and\ \bibinfo {author} {\bibfnamefont {University}\ \bibnamefont
  {of~California~Press}},\ }\href
  {https://books.google.co.uk/books?id=avgkDQAAQBAJ} {\emph {\bibinfo {title}
  {The Theory of Atomic Structure and Spectra}}},\ Los Alamos Series in Basic
  and Applied Sciences\ (\bibinfo  {publisher} {University of California
  Press},\ \bibinfo {year} {1981})\BibitemShut {NoStop}%
\bibitem [{\citenamefont {Zheludev}\ and\ \citenamefont
  {Kivshar}(2012)}]{Zheludev12}%
  \BibitemOpen
  \bibfield  {author} {\bibinfo {author} {\bibfnamefont {Nikolay~I.}\
  \bibnamefont {Zheludev}}\ and\ \bibinfo {author} {\bibfnamefont {Yuri~S.}\
  \bibnamefont {Kivshar}},\ }\bibfield  {title} {\enquote {\bibinfo {title}
  {From metamaterials to metadevices},}\ }\href {\doibase 10.1038/nmat3431}
  {\bibfield  {journal} {\bibinfo  {journal} {Nature Materials}\ }\textbf
  {\bibinfo {volume} {11}},\ \bibinfo {pages} {917--924} (\bibinfo {year}
  {2012})}\BibitemShut {NoStop}%
\bibitem [{\citenamefont {Yu}\ and\ \citenamefont
  {Capasso}(2014)}]{capasso_review}%
  \BibitemOpen
  \bibfield  {author} {\bibinfo {author} {\bibfnamefont {Nanfang}\ \bibnamefont
  {Yu}}\ and\ \bibinfo {author} {\bibfnamefont {Federico}\ \bibnamefont
  {Capasso}},\ }\bibfield  {title} {\enquote {\bibinfo {title} {Flat optics
  with designer metasurfaces},}\ }\href {\doibase 10.1038/nmat3839} {\bibfield
  {journal} {\bibinfo  {journal} {Nature Materials}\ }\textbf {\bibinfo
  {volume} {13}},\ \bibinfo {pages} {139--150} (\bibinfo {year}
  {2014})}\BibitemShut {NoStop}%
\bibitem [{\citenamefont {Novotny}\ and\ \citenamefont
  {Hecht}(2012)}]{Novotny12}%
  \BibitemOpen
  \bibfield  {author} {\bibinfo {author} {\bibfnamefont {L.}~\bibnamefont
  {Novotny}}\ and\ \bibinfo {author} {\bibfnamefont {B.}~\bibnamefont
  {Hecht}},\ }\href {https://books.google.co.uk/books?id=FUAAxtxvGI4C} {\emph
  {\bibinfo {title} {Principles of Nano-Optics}}},\ Principles of Nano-optics\
  (\bibinfo  {publisher} {Cambridge University Press},\ \bibinfo {year}
  {2012})\BibitemShut {NoStop}%
\bibitem [{\citenamefont {Pendry}(2000)}]{PendryPRL2000}%
  \BibitemOpen
  \bibfield  {author} {\bibinfo {author} {\bibfnamefont {J.~B.}\ \bibnamefont
  {Pendry}},\ }\bibfield  {title} {\enquote {\bibinfo {title} {Negative
  refraction makes a perfect lens},}\ }\href@noop {} {\bibfield  {journal}
  {\bibinfo  {journal} {Phys. Rev. Lett.}\ }\textbf {\bibinfo {volume} {85}},\
  \bibinfo {pages} {3966} (\bibinfo {year} {2000})}\BibitemShut {NoStop}%
\bibitem [{\citenamefont {Pendry}\ \emph {et~al.}(2006)\citenamefont {Pendry},
  \citenamefont {Schurig},\ and\ \citenamefont {Smith}}]{PendryEtAlSCI2006}%
  \BibitemOpen
  \bibfield  {author} {\bibinfo {author} {\bibfnamefont {J.~B.}\ \bibnamefont
  {Pendry}}, \bibinfo {author} {\bibfnamefont {D.}~\bibnamefont {Schurig}}, \
  and\ \bibinfo {author} {\bibfnamefont {D.~R.}\ \bibnamefont {Smith}},\
  }\bibfield  {title} {\enquote {\bibinfo {title} {Controlling electromagnetic
  fields},}\ }\href@noop {} {\bibfield  {journal} {\bibinfo  {journal}
  {Science}\ }\textbf {\bibinfo {volume} {312}},\ \bibinfo {pages} {1780}
  (\bibinfo {year} {2006})}\BibitemShut {NoStop}%
\bibitem [{\citenamefont {Leonhardt}(2006)}]{LeonhardtSCI2006}%
  \BibitemOpen
  \bibfield  {author} {\bibinfo {author} {\bibfnamefont {Ulf}\ \bibnamefont
  {Leonhardt}},\ }\bibfield  {title} {\enquote {\bibinfo {title} {Optical
  conformal mapping},}\ }\href@noop {} {\bibfield  {journal} {\bibinfo
  {journal} {Science}\ }\textbf {\bibinfo {volume} {312}},\ \bibinfo {pages}
  {1777} (\bibinfo {year} {2006})}\BibitemShut {NoStop}%
\bibitem [{\citenamefont {Schurig}\ \emph {et~al.}(2006)\citenamefont
  {Schurig}, \citenamefont {Mock}, \citenamefont {Justice}, \citenamefont
  {Cummer}, \citenamefont {Pendry}, \citenamefont {Starr},\ and\ \citenamefont
  {Smith}}]{SchurigSci2006}%
  \BibitemOpen
  \bibfield  {author} {\bibinfo {author} {\bibfnamefont {D.}~\bibnamefont
  {Schurig}}, \bibinfo {author} {\bibfnamefont {J.~J.}\ \bibnamefont {Mock}},
  \bibinfo {author} {\bibfnamefont {B.~J.}\ \bibnamefont {Justice}}, \bibinfo
  {author} {\bibfnamefont {S.~A.}\ \bibnamefont {Cummer}}, \bibinfo {author}
  {\bibfnamefont {J.~B.}\ \bibnamefont {Pendry}}, \bibinfo {author}
  {\bibfnamefont {A.~F.}\ \bibnamefont {Starr}}, \ and\ \bibinfo {author}
  {\bibfnamefont {D.~R.}\ \bibnamefont {Smith}},\ }\bibfield  {title} {\enquote
  {\bibinfo {title} {Metamaterial electromagnetic cloak at microwave
  frequencies},}\ }\href@noop {} {\bibfield  {journal} {\bibinfo  {journal}
  {Science}\ }\textbf {\bibinfo {volume} {314}},\ \bibinfo {pages} {977}
  (\bibinfo {year} {2006})}\BibitemShut {NoStop}%
\bibitem [{\citenamefont {Huygens}(1690)}]{Huygens}%
  \BibitemOpen
  \bibfield  {author} {\bibinfo {author} {\bibfnamefont {C.}~\bibnamefont
  {Huygens}},\ }\href@noop {} {\enquote {\bibinfo {title} {Trait{\'e} de la
  lumi{\'e}re},}\ } (\bibinfo {year} {1690})\BibitemShut {NoStop}%
\bibitem [{\citenamefont {Love}(1901)}]{Love1901}%
  \BibitemOpen
  \bibfield  {author} {\bibinfo {author} {\bibfnamefont {A.~E.~H.}\
  \bibnamefont {Love}},\ }\bibfield  {title} {\enquote {\bibinfo {title} {The
  integration of the equations of propagation of electric waves},}\ }\href
  {\doibase 10.1098/rsta.1901.0013} {\bibfield  {journal} {\bibinfo  {journal}
  {Phil. Trans. R. Soc. London A}\ }\textbf {\bibinfo {volume} {197}},\
  \bibinfo {pages} {1} (\bibinfo {year} {1901})}\BibitemShut {NoStop}%
\bibitem [{\citenamefont {Weitenberg}\ \emph {et~al.}(2011)\citenamefont
  {Weitenberg}, \citenamefont {Endres}, \citenamefont {Sherson}, \citenamefont
  {Cheneau}, \citenamefont {Schau{\ss}}, \citenamefont {Fukuhara},
  \citenamefont {Bloch},\ and\ \citenamefont {Kuhr}}]{Weitenberg11}%
  \BibitemOpen
  \bibfield  {author} {\bibinfo {author} {\bibfnamefont {Christof}\
  \bibnamefont {Weitenberg}}, \bibinfo {author} {\bibfnamefont {Manuel}\
  \bibnamefont {Endres}}, \bibinfo {author} {\bibfnamefont {Jacob~F.}\
  \bibnamefont {Sherson}}, \bibinfo {author} {\bibfnamefont {Marc}\
  \bibnamefont {Cheneau}}, \bibinfo {author} {\bibfnamefont {Peter}\
  \bibnamefont {Schau{\ss}}}, \bibinfo {author} {\bibfnamefont {Takeshi}\
  \bibnamefont {Fukuhara}}, \bibinfo {author} {\bibfnamefont {Immanuel}\
  \bibnamefont {Bloch}}, \ and\ \bibinfo {author} {\bibfnamefont {Stefan}\
  \bibnamefont {Kuhr}},\ }\bibfield  {title} {\enquote {\bibinfo {title}
  {Single-spin addressing in an atomic mott insulator},}\ }\href {\doibase
  10.1038/nature09827} {\bibfield  {journal} {\bibinfo  {journal} {Nature}\
  }\textbf {\bibinfo {volume} {471}},\ \bibinfo {pages} {319--324} (\bibinfo
  {year} {2011})}\BibitemShut {NoStop}%
\bibitem [{\citenamefont {Lester}\ \emph {et~al.}(2015)\citenamefont {Lester},
  \citenamefont {Luick}, \citenamefont {Kaufman}, \citenamefont {Reynolds},\
  and\ \citenamefont {Regal}}]{Lester15}%
  \BibitemOpen
  \bibfield  {author} {\bibinfo {author} {\bibfnamefont {Brian~J.}\
  \bibnamefont {Lester}}, \bibinfo {author} {\bibfnamefont {Niclas}\
  \bibnamefont {Luick}}, \bibinfo {author} {\bibfnamefont {Adam~M.}\
  \bibnamefont {Kaufman}}, \bibinfo {author} {\bibfnamefont {Collin~M.}\
  \bibnamefont {Reynolds}}, \ and\ \bibinfo {author} {\bibfnamefont {Cindy~A.}\
  \bibnamefont {Regal}},\ }\bibfield  {title} {\enquote {\bibinfo {title}
  {Rapid production of uniformly filled arrays of neutral atoms},}\ }\href
  {\doibase 10.1103/PhysRevLett.115.073003} {\bibfield  {journal} {\bibinfo
  {journal} {Phys. Rev. Lett.}\ }\textbf {\bibinfo {volume} {115}},\ \bibinfo
  {pages} {073003} (\bibinfo {year} {2015})}\BibitemShut {NoStop}%
\bibitem [{\citenamefont {Xia}\ \emph {et~al.}(2015)\citenamefont {Xia},
  \citenamefont {Lichtman}, \citenamefont {Maller}, \citenamefont {Carr},
  \citenamefont {Piotrowicz}, \citenamefont {Isenhower},\ and\ \citenamefont
  {Saffman}}]{Xia15}%
  \BibitemOpen
  \bibfield  {author} {\bibinfo {author} {\bibfnamefont {T.}~\bibnamefont
  {Xia}}, \bibinfo {author} {\bibfnamefont {M.}~\bibnamefont {Lichtman}},
  \bibinfo {author} {\bibfnamefont {K.}~\bibnamefont {Maller}}, \bibinfo
  {author} {\bibfnamefont {A.~W.}\ \bibnamefont {Carr}}, \bibinfo {author}
  {\bibfnamefont {M.~J.}\ \bibnamefont {Piotrowicz}}, \bibinfo {author}
  {\bibfnamefont {L.}~\bibnamefont {Isenhower}}, \ and\ \bibinfo {author}
  {\bibfnamefont {M.}~\bibnamefont {Saffman}},\ }\bibfield  {title} {\enquote
  {\bibinfo {title} {Randomized benchmarking of single-qubit gates in a 2d
  array of neutral-atom qubits},}\ }\href {\doibase
  10.1103/PhysRevLett.114.100503} {\bibfield  {journal} {\bibinfo  {journal}
  {Phys. Rev. Lett.}\ }\textbf {\bibinfo {volume} {114}},\ \bibinfo {pages}
  {100503} (\bibinfo {year} {2015})}\BibitemShut {NoStop}%
\bibitem [{\citenamefont {Endres}\ \emph {et~al.}(2016)\citenamefont {Endres},
  \citenamefont {Bernien}, \citenamefont {Keesling}, \citenamefont {Levine},
  \citenamefont {Anschuetz}, \citenamefont {Krajenbrink}, \citenamefont
  {Senko}, \citenamefont {Vuletic}, \citenamefont {Greiner},\ and\
  \citenamefont {Lukin}}]{Endres16}%
  \BibitemOpen
  \bibfield  {author} {\bibinfo {author} {\bibfnamefont {Manuel}\ \bibnamefont
  {Endres}}, \bibinfo {author} {\bibfnamefont {Hannes}\ \bibnamefont
  {Bernien}}, \bibinfo {author} {\bibfnamefont {Alexander}\ \bibnamefont
  {Keesling}}, \bibinfo {author} {\bibfnamefont {Harry}\ \bibnamefont
  {Levine}}, \bibinfo {author} {\bibfnamefont {Eric~R.}\ \bibnamefont
  {Anschuetz}}, \bibinfo {author} {\bibfnamefont {Alexandre}\ \bibnamefont
  {Krajenbrink}}, \bibinfo {author} {\bibfnamefont {Crystal}\ \bibnamefont
  {Senko}}, \bibinfo {author} {\bibfnamefont {Vladan}\ \bibnamefont {Vuletic}},
  \bibinfo {author} {\bibfnamefont {Markus}\ \bibnamefont {Greiner}}, \ and\
  \bibinfo {author} {\bibfnamefont {Mikhail~D.}\ \bibnamefont {Lukin}},\
  }\bibfield  {title} {\enquote {\bibinfo {title} {Atom-by-atom assembly of
  defect-free one-dimensional cold atom arrays},}\ }\href {\doibase
  10.1126/science.aah3752} {\bibfield  {journal} {\bibinfo  {journal}
  {Science}\ }\textbf {\bibinfo {volume} {354}},\ \bibinfo {pages} {1024--1027}
  (\bibinfo {year} {2016})}\BibitemShut {NoStop}%
\bibitem [{\citenamefont {Barredo}\ \emph {et~al.}(2016)\citenamefont
  {Barredo}, \citenamefont {de~L{\'e}s{\'e}leuc}, \citenamefont {Lienhard},
  \citenamefont {Lahaye},\ and\ \citenamefont {Browaeys}}]{Barredo16}%
  \BibitemOpen
  \bibfield  {author} {\bibinfo {author} {\bibfnamefont {Daniel}\ \bibnamefont
  {Barredo}}, \bibinfo {author} {\bibfnamefont {Sylvain}\ \bibnamefont
  {de~L{\'e}s{\'e}leuc}}, \bibinfo {author} {\bibfnamefont {Vincent}\
  \bibnamefont {Lienhard}}, \bibinfo {author} {\bibfnamefont {Thierry}\
  \bibnamefont {Lahaye}}, \ and\ \bibinfo {author} {\bibfnamefont {Antoine}\
  \bibnamefont {Browaeys}},\ }\bibfield  {title} {\enquote {\bibinfo {title}
  {An atom-by-atom assembler of defect-free arbitrary two-dimensional atomic
  arrays},}\ }\href {\doibase 10.1126/science.aah3778} {\bibfield  {journal}
  {\bibinfo  {journal} {Science}\ }\textbf {\bibinfo {volume} {354}},\ \bibinfo
  {pages} {1021--1023} (\bibinfo {year} {2016})}\BibitemShut {NoStop}%
\bibitem [{\citenamefont {Kim}\ \emph {et~al.}(2016)\citenamefont {Kim},
  \citenamefont {Lee}, \citenamefont {Lee}, \citenamefont {Jo}, \citenamefont
  {Song},\ and\ \citenamefont {Ahn}}]{Kim16}%
  \BibitemOpen
  \bibfield  {author} {\bibinfo {author} {\bibfnamefont {Hyosub}\ \bibnamefont
  {Kim}}, \bibinfo {author} {\bibfnamefont {Woojun}\ \bibnamefont {Lee}},
  \bibinfo {author} {\bibfnamefont {Han-gyeol}\ \bibnamefont {Lee}}, \bibinfo
  {author} {\bibfnamefont {Hanlae}\ \bibnamefont {Jo}}, \bibinfo {author}
  {\bibfnamefont {Yunheung}\ \bibnamefont {Song}}, \ and\ \bibinfo {author}
  {\bibfnamefont {Jaewook}\ \bibnamefont {Ahn}},\ }\bibfield  {title} {\enquote
  {\bibinfo {title} {In situ single-atom array synthesis using dynamic
  holographic optical tweezers},}\ }\href {\doibase 10.1038/ncomms13317}
  {\bibfield  {journal} {\bibinfo  {journal} {Nature Communications}\ }\textbf
  {\bibinfo {volume} {7}},\ \bibinfo {pages} {13317} (\bibinfo {year}
  {2016})}\BibitemShut {NoStop}%
\bibitem [{\citenamefont {Cooper}\ \emph {et~al.}(2018)\citenamefont {Cooper},
  \citenamefont {Covey}, \citenamefont {Madjarov}, \citenamefont {Porsev},
  \citenamefont {Safronova},\ and\ \citenamefont {Endres}}]{Cooper18}%
  \BibitemOpen
  \bibfield  {author} {\bibinfo {author} {\bibfnamefont {Alexandre}\
  \bibnamefont {Cooper}}, \bibinfo {author} {\bibfnamefont {Jacob~P.}\
  \bibnamefont {Covey}}, \bibinfo {author} {\bibfnamefont {Ivaylo~S.}\
  \bibnamefont {Madjarov}}, \bibinfo {author} {\bibfnamefont {Sergey~G.}\
  \bibnamefont {Porsev}}, \bibinfo {author} {\bibfnamefont {Marianna~S.}\
  \bibnamefont {Safronova}}, \ and\ \bibinfo {author} {\bibfnamefont {Manuel}\
  \bibnamefont {Endres}},\ }\bibfield  {title} {\enquote {\bibinfo {title}
  {Alkaline-earth atoms in optical tweezers},}\ }\href {\doibase
  10.1103/PhysRevX.8.041055} {\bibfield  {journal} {\bibinfo  {journal} {Phys.
  Rev. X}\ }\textbf {\bibinfo {volume} {8}},\ \bibinfo {pages} {041055}
  (\bibinfo {year} {2018})}\BibitemShut {NoStop}%
\bibitem [{\citenamefont {Jenkins}\ and\ \citenamefont
  {Ruostekoski}(2012)}]{Jenkins2012a}%
  \BibitemOpen
  \bibfield  {author} {\bibinfo {author} {\bibfnamefont {Stewart~D.}\
  \bibnamefont {Jenkins}}\ and\ \bibinfo {author} {\bibfnamefont {Janne}\
  \bibnamefont {Ruostekoski}},\ }\bibfield  {title} {\enquote {\bibinfo {title}
  {Controlled manipulation of light by cooperative response of atoms in an
  optical lattice},}\ }\href {\doibase 10.1103/PhysRevA.86.031602} {\bibfield
  {journal} {\bibinfo  {journal} {Phys. Rev. A}\ }\textbf {\bibinfo {volume}
  {86}},\ \bibinfo {pages} {031602(R)} (\bibinfo {year} {2012})}\BibitemShut
  {NoStop}%
\bibitem [{\citenamefont {Perczel}\ \emph {et~al.}(2017)\citenamefont
  {Perczel}, \citenamefont {Borregaard}, \citenamefont {Chang}, \citenamefont
  {Pichler}, \citenamefont {Yelin}, \citenamefont {Zoller},\ and\ \citenamefont
  {Lukin}}]{Perczel2017a}%
  \BibitemOpen
  \bibfield  {author} {\bibinfo {author} {\bibfnamefont {Janos}\ \bibnamefont
  {Perczel}}, \bibinfo {author} {\bibfnamefont {Johannes}\ \bibnamefont
  {Borregaard}}, \bibinfo {author} {\bibfnamefont {Darrick~E}\ \bibnamefont
  {Chang}}, \bibinfo {author} {\bibfnamefont {Hannes}\ \bibnamefont {Pichler}},
  \bibinfo {author} {\bibfnamefont {Susanne~F}\ \bibnamefont {Yelin}}, \bibinfo
  {author} {\bibfnamefont {Peter}\ \bibnamefont {Zoller}}, \ and\ \bibinfo
  {author} {\bibfnamefont {Mikhail~D.}\ \bibnamefont {Lukin}},\ }\bibfield
  {title} {\enquote {\bibinfo {title} {{Photonic band structure of
  two-dimensional atomic lattices}},}\ }\href {\doibase
  10.1103/PhysRevA.96.063801} {\bibfield  {journal} {\bibinfo  {journal} {Phys.
  Rev. A}\ }\textbf {\bibinfo {volume} {96}},\ \bibinfo {pages} {063801}
  (\bibinfo {year} {2017})}\BibitemShut {NoStop}%
\bibitem [{\citenamefont {Bettles}\ \emph {et~al.}(2017)\citenamefont
  {Bettles}, \citenamefont {Min\'{a}\v{r}}, \citenamefont {Adams},
  \citenamefont {Lesanovsky},\ and\ \citenamefont {Olmos}}]{Bettles2017}%
  \BibitemOpen
  \bibfield  {author} {\bibinfo {author} {\bibfnamefont {Robert~J.}\
  \bibnamefont {Bettles}}, \bibinfo {author} {\bibfnamefont {Ji\v{r}\'{i}}\
  \bibnamefont {Min\'{a}\v{r}}}, \bibinfo {author} {\bibfnamefont {Charles~S.}\
  \bibnamefont {Adams}}, \bibinfo {author} {\bibfnamefont {Igor}\ \bibnamefont
  {Lesanovsky}}, \ and\ \bibinfo {author} {\bibfnamefont {Beatriz}\
  \bibnamefont {Olmos}},\ }\bibfield  {title} {\enquote {\bibinfo {title}
  {{Topological properties of a dense atomic lattice gas}},}\ }\href {\doibase
  10.1103/PhysRevA.96.041603} {\bibfield  {journal} {\bibinfo  {journal} {Phys.
  Rev. A}\ }\textbf {\bibinfo {volume} {96}},\ \bibinfo {pages} {041603(R)}
  (\bibinfo {year} {2017})}\BibitemShut {NoStop}%
\bibitem [{\citenamefont {Bettles}\ \emph {et~al.}(2016)\citenamefont
  {Bettles}, \citenamefont {Gardiner},\ and\ \citenamefont
  {Adams}}]{Bettles2016}%
  \BibitemOpen
  \bibfield  {author} {\bibinfo {author} {\bibfnamefont {Robert~J.}\
  \bibnamefont {Bettles}}, \bibinfo {author} {\bibfnamefont {S.~A.}\
  \bibnamefont {Gardiner}}, \ and\ \bibinfo {author} {\bibfnamefont
  {Charles~S.}\ \bibnamefont {Adams}},\ }\bibfield  {title} {\enquote {\bibinfo
  {title} {{Enhanced Optical Cross Section via Collective Coupling of Atomic
  Dipoles in a 2D Array}},}\ }\href {\doibase 10.1103/PhysRevLett.116.103602}
  {\bibfield  {journal} {\bibinfo  {journal} {Phys. Rev. Lett.}\ }\textbf
  {\bibinfo {volume} {116}},\ \bibinfo {pages} {103602} (\bibinfo {year}
  {2016})}\BibitemShut {NoStop}%
\bibitem [{\citenamefont {Facchinetti}\ \emph {et~al.}(2016)\citenamefont
  {Facchinetti}, \citenamefont {Jenkins},\ and\ \citenamefont
  {Ruostekoski}}]{Facchinetti16}%
  \BibitemOpen
  \bibfield  {author} {\bibinfo {author} {\bibfnamefont {G.}~\bibnamefont
  {Facchinetti}}, \bibinfo {author} {\bibfnamefont {S.~D.}\ \bibnamefont
  {Jenkins}}, \ and\ \bibinfo {author} {\bibfnamefont {J.}~\bibnamefont
  {Ruostekoski}},\ }\bibfield  {title} {\enquote {\bibinfo {title} {Storing
  light with subradiant correlations in arrays of atoms},}\ }\href {\doibase
  10.1103/PhysRevLett.117.243601} {\bibfield  {journal} {\bibinfo  {journal}
  {Phys. Rev. Lett.}\ }\textbf {\bibinfo {volume} {117}},\ \bibinfo {pages}
  {243601} (\bibinfo {year} {2016})}\BibitemShut {NoStop}%
\bibitem [{\citenamefont {Facchinetti}\ and\ \citenamefont
  {Ruostekoski}(2018)}]{Facchinetti18}%
  \BibitemOpen
  \bibfield  {author} {\bibinfo {author} {\bibfnamefont {G.}~\bibnamefont
  {Facchinetti}}\ and\ \bibinfo {author} {\bibfnamefont {J.}~\bibnamefont
  {Ruostekoski}},\ }\bibfield  {title} {\enquote {\bibinfo {title} {Interaction
  of light with planar lattices of atoms: Reflection, transmission, and
  cooperative magnetometry},}\ }\href {\doibase 10.1103/PhysRevA.97.023833}
  {\bibfield  {journal} {\bibinfo  {journal} {Phys. Rev. A}\ }\textbf {\bibinfo
  {volume} {97}},\ \bibinfo {pages} {023833} (\bibinfo {year}
  {2018})}\BibitemShut {NoStop}%
\bibitem [{\citenamefont {Shahmoon}\ \emph {et~al.}(2017)\citenamefont
  {Shahmoon}, \citenamefont {Wild}, \citenamefont {Lukin},\ and\ \citenamefont
  {Yelin}}]{Shahmoon}%
  \BibitemOpen
  \bibfield  {author} {\bibinfo {author} {\bibfnamefont {Ephraim}\ \bibnamefont
  {Shahmoon}}, \bibinfo {author} {\bibfnamefont {Dominik~S.}\ \bibnamefont
  {Wild}}, \bibinfo {author} {\bibfnamefont {Mikhail~D.}\ \bibnamefont
  {Lukin}}, \ and\ \bibinfo {author} {\bibfnamefont {Susanne~F.}\ \bibnamefont
  {Yelin}},\ }\bibfield  {title} {\enquote {\bibinfo {title} {Cooperative
  resonances in light scattering from two-dimensional atomic arrays},}\ }\href
  {\doibase 10.1103/PhysRevLett.118.113601} {\bibfield  {journal} {\bibinfo
  {journal} {Phys. Rev. Lett.}\ }\textbf {\bibinfo {volume} {118}},\ \bibinfo
  {pages} {113601} (\bibinfo {year} {2017})}\BibitemShut {NoStop}%
\bibitem [{\citenamefont {Plankensteiner}\ \emph {et~al.}(2017)\citenamefont
  {Plankensteiner}, \citenamefont {Sommer}, \citenamefont {Ritsch},\ and\
  \citenamefont {Genes}}]{Plankensteiner2017}%
  \BibitemOpen
  \bibfield  {author} {\bibinfo {author} {\bibfnamefont {David}\ \bibnamefont
  {Plankensteiner}}, \bibinfo {author} {\bibfnamefont {Christian}\ \bibnamefont
  {Sommer}}, \bibinfo {author} {\bibfnamefont {Helmut}\ \bibnamefont {Ritsch}},
  \ and\ \bibinfo {author} {\bibfnamefont {Claudiu}\ \bibnamefont {Genes}},\
  }\bibfield  {title} {\enquote {\bibinfo {title} {{Cavity Antiresonance
  Spectroscopy of Dipole Coupled Subradiant Arrays}},}\ }\href {\doibase
  10.1103/PhysRevLett.119.093601} {\bibfield  {journal} {\bibinfo  {journal}
  {Phys. Rev. Lett.}\ }\textbf {\bibinfo {volume} {119}},\ \bibinfo {pages}
  {093601} (\bibinfo {year} {2017})}\BibitemShut {NoStop}%
\bibitem [{\citenamefont {Asenjo-Garcia}\ \emph {et~al.}(2017)\citenamefont
  {Asenjo-Garcia}, \citenamefont {Moreno-Cardoner}, \citenamefont {Albrecht},
  \citenamefont {Kimble},\ and\ \citenamefont {Chang}}]{Asenjo-Garcia2017a}%
  \BibitemOpen
  \bibfield  {author} {\bibinfo {author} {\bibfnamefont {Ana}\ \bibnamefont
  {Asenjo-Garcia}}, \bibinfo {author} {\bibfnamefont {M.}~\bibnamefont
  {Moreno-Cardoner}}, \bibinfo {author} {\bibfnamefont {A.}~\bibnamefont
  {Albrecht}}, \bibinfo {author} {\bibfnamefont {H.~J.}\ \bibnamefont
  {Kimble}}, \ and\ \bibinfo {author} {\bibfnamefont {D.~E.}\ \bibnamefont
  {Chang}},\ }\bibfield  {title} {\enquote {\bibinfo {title} {{Exponential
  Improvement in Photon Storage Fidelities Using Subradiance and “Selective
  Radiance” in Atomic Arrays}},}\ }\href {\doibase 10.1103/PhysRevX.7.031024}
  {\bibfield  {journal} {\bibinfo  {journal} {Phys. Rev. X}\ }\textbf {\bibinfo
  {volume} {7}},\ \bibinfo {pages} {031024} (\bibinfo {year}
  {2017})}\BibitemShut {NoStop}%
\bibitem [{\citenamefont {Jen}(2017)}]{Jen17}%
  \BibitemOpen
  \bibfield  {author} {\bibinfo {author} {\bibfnamefont {H.~H.}\ \bibnamefont
  {Jen}},\ }\bibfield  {title} {\enquote {\bibinfo {title} {Phase-imprinted
  multiphoton subradiant states},}\ }\href {\doibase
  10.1103/PhysRevA.96.023814} {\bibfield  {journal} {\bibinfo  {journal} {Phys.
  Rev. A}\ }\textbf {\bibinfo {volume} {96}},\ \bibinfo {pages} {023814}
  (\bibinfo {year} {2017})}\BibitemShut {NoStop}%
\bibitem [{\citenamefont {Guimond}\ \emph {et~al.}(2019)\citenamefont
  {Guimond}, \citenamefont {Grankin}, \citenamefont {Vasilyev}, \citenamefont
  {Vermersch},\ and\ \citenamefont {Zoller}}]{Guimond2019}%
  \BibitemOpen
  \bibfield  {author} {\bibinfo {author} {\bibfnamefont {P.-O.}\ \bibnamefont
  {Guimond}}, \bibinfo {author} {\bibfnamefont {A.}~\bibnamefont {Grankin}},
  \bibinfo {author} {\bibfnamefont {D.~V.}\ \bibnamefont {Vasilyev}}, \bibinfo
  {author} {\bibfnamefont {B.}~\bibnamefont {Vermersch}}, \ and\ \bibinfo
  {author} {\bibfnamefont {P.}~\bibnamefont {Zoller}},\ }\bibfield  {title}
  {\enquote {\bibinfo {title} {Subradiant bell states in distant atomic
  arrays},}\ }\href {\doibase 10.1103/PhysRevLett.122.093601} {\bibfield
  {journal} {\bibinfo  {journal} {Phys. Rev. Lett.}\ }\textbf {\bibinfo
  {volume} {122}},\ \bibinfo {pages} {093601} (\bibinfo {year}
  {2019})}\BibitemShut {NoStop}%
\bibitem [{\citenamefont {Hebenstreit}\ \emph {et~al.}(2017)\citenamefont
  {Hebenstreit}, \citenamefont {Kraus}, \citenamefont {Ostermann},\ and\
  \citenamefont {Ritsch}}]{Ritsch_subr}%
  \BibitemOpen
  \bibfield  {author} {\bibinfo {author} {\bibfnamefont {Martin}\ \bibnamefont
  {Hebenstreit}}, \bibinfo {author} {\bibfnamefont {Barbara}\ \bibnamefont
  {Kraus}}, \bibinfo {author} {\bibfnamefont {Laurin}\ \bibnamefont
  {Ostermann}}, \ and\ \bibinfo {author} {\bibfnamefont {Helmut}\ \bibnamefont
  {Ritsch}},\ }\bibfield  {title} {\enquote {\bibinfo {title} {Subradiance via
  entanglement in atoms with several independent decay channels},}\ }\href
  {\doibase 10.1103/PhysRevLett.118.143602} {\bibfield  {journal} {\bibinfo
  {journal} {Phys. Rev. Lett.}\ }\textbf {\bibinfo {volume} {118}},\ \bibinfo
  {pages} {143602} (\bibinfo {year} {2017})}\BibitemShut {NoStop}%
\bibitem [{\citenamefont {Kr{\"{a}}mer}\ \emph {et~al.}(2016)\citenamefont
  {Kr{\"{a}}mer}, \citenamefont {Ostermann},\ and\ \citenamefont
  {Ritsch}}]{Kramer2016}%
  \BibitemOpen
  \bibfield  {author} {\bibinfo {author} {\bibfnamefont {Sebastian}\
  \bibnamefont {Kr{\"{a}}mer}}, \bibinfo {author} {\bibfnamefont {Laurin}\
  \bibnamefont {Ostermann}}, \ and\ \bibinfo {author} {\bibfnamefont {Helmut}\
  \bibnamefont {Ritsch}},\ }\bibfield  {title} {\enquote {\bibinfo {title}
  {{Optimized geometries for future generation optical lattice clocks}},}\
  }\href {\doibase 10.1209/0295-5075/114/14003} {\bibfield  {journal} {\bibinfo
   {journal} {Europhys. Lett.}\ }\textbf {\bibinfo {volume} {114}},\ \bibinfo
  {pages} {14003} (\bibinfo {year} {2016})}\BibitemShut {NoStop}%
\bibitem [{\citenamefont {Sutherland}\ and\ \citenamefont
  {Robicheaux}(2016{\natexlab{a}})}]{Sutherland1D}%
  \BibitemOpen
  \bibfield  {author} {\bibinfo {author} {\bibfnamefont {R.~T.}\ \bibnamefont
  {Sutherland}}\ and\ \bibinfo {author} {\bibfnamefont {F.}~\bibnamefont
  {Robicheaux}},\ }\bibfield  {title} {\enquote {\bibinfo {title} {Collective
  dipole-dipole interactions in an atomic array},}\ }\href {\doibase
  10.1103/PhysRevA.94.013847} {\bibfield  {journal} {\bibinfo  {journal} {Phys.
  Rev. A}\ }\textbf {\bibinfo {volume} {94}},\ \bibinfo {pages} {013847}
  (\bibinfo {year} {2016}{\natexlab{a}})}\BibitemShut {NoStop}%
\bibitem [{\citenamefont {Yoo}\ and\ \citenamefont {Paik}(2016)}]{Yoo2016}%
  \BibitemOpen
  \bibfield  {author} {\bibinfo {author} {\bibfnamefont {Sung-Mi}\ \bibnamefont
  {Yoo}}\ and\ \bibinfo {author} {\bibfnamefont {Sun~Mok}\ \bibnamefont
  {Paik}},\ }\bibfield  {title} {\enquote {\bibinfo {title} {{Cooperative
  optical response of 2D dense lattices with strongly correlated dipoles}},}\
  }\href {\doibase 10.1364/OE.24.002156} {\bibfield  {journal} {\bibinfo
  {journal} {Opt. Express}\ }\textbf {\bibinfo {volume} {24}},\ \bibinfo
  {pages} {2156} (\bibinfo {year} {2016})}\BibitemShut {NoStop}%
\bibitem [{\citenamefont {Zhang}\ and\ \citenamefont
  {M\o{}lmer}(2019)}]{Zhang2018}%
  \BibitemOpen
  \bibfield  {author} {\bibinfo {author} {\bibfnamefont {Yu-Xiang}\
  \bibnamefont {Zhang}}\ and\ \bibinfo {author} {\bibfnamefont {Klaus}\
  \bibnamefont {M\o{}lmer}},\ }\bibfield  {title} {\enquote {\bibinfo {title}
  {Theory of subradiant states of a one-dimensional two-level atom chain},}\
  }\href {\doibase 10.1103/PhysRevLett.122.203605} {\bibfield  {journal}
  {\bibinfo  {journal} {Phys. Rev. Lett.}\ }\textbf {\bibinfo {volume} {122}},\
  \bibinfo {pages} {203605} (\bibinfo {year} {2019})}\BibitemShut {NoStop}%
\bibitem [{\citenamefont {Mkhitaryan}\ \emph {et~al.}(2018)\citenamefont
  {Mkhitaryan}, \citenamefont {Meng}, \citenamefont {Marini},\ and\
  \citenamefont {de~Abajo}}]{Mkhitaryan18}%
  \BibitemOpen
  \bibfield  {author} {\bibinfo {author} {\bibfnamefont {Vahagn}\ \bibnamefont
  {Mkhitaryan}}, \bibinfo {author} {\bibfnamefont {Lijun}\ \bibnamefont
  {Meng}}, \bibinfo {author} {\bibfnamefont {Andrea}\ \bibnamefont {Marini}}, \
  and\ \bibinfo {author} {\bibfnamefont {F.~J.}\ \bibnamefont
  {de~Abajo}},\ }\bibfield  {title} {\enquote {\bibinfo {title} {Lasing and
  amplification from two-dimensional atom arrays},}\ }\href {\doibase
  10.1103/PhysRevLett.121.163602} {\bibfield  {journal} {\bibinfo  {journal}
  {Phys. Rev. Lett.}\ }\textbf {\bibinfo {volume} {121}},\ \bibinfo {pages}
  {163602} (\bibinfo {year} {2018})}\BibitemShut {NoStop}%
\bibitem [{\citenamefont {Bhatti}\ \emph {et~al.}(2018)\citenamefont {Bhatti},
  \citenamefont {Schneider}, \citenamefont {Oppel},\ and\ \citenamefont {von
  Zanthier}}]{Bhatti18}%
  \BibitemOpen
  \bibfield  {author} {\bibinfo {author} {\bibfnamefont {Daniel}\ \bibnamefont
  {Bhatti}}, \bibinfo {author} {\bibfnamefont {Raimund}\ \bibnamefont
  {Schneider}}, \bibinfo {author} {\bibfnamefont {Steffen}\ \bibnamefont
  {Oppel}}, \ and\ \bibinfo {author} {\bibfnamefont {Joachim}\ \bibnamefont
  {von Zanthier}},\ }\bibfield  {title} {\enquote {\bibinfo {title}
  {Directional dicke subradiance with nonclassical and classical light
  sources},}\ }\href {\doibase 10.1103/PhysRevLett.120.113603} {\bibfield
  {journal} {\bibinfo  {journal} {Phys. Rev. Lett.}\ }\textbf {\bibinfo
  {volume} {120}},\ \bibinfo {pages} {113603} (\bibinfo {year}
  {2018})}\BibitemShut {NoStop}%
\bibitem [{\citenamefont {Henriet}\ \emph {et~al.}(2019)\citenamefont
  {Henriet}, \citenamefont {Douglas}, \citenamefont {Chang},\ and\
  \citenamefont {Albrecht}}]{Henriet2018}%
  \BibitemOpen
  \bibfield  {author} {\bibinfo {author} {\bibfnamefont {Lo\"{\i}c}\
  \bibnamefont {Henriet}}, \bibinfo {author} {\bibfnamefont {James~S.}\
  \bibnamefont {Douglas}}, \bibinfo {author} {\bibfnamefont {Darrick~E.}\
  \bibnamefont {Chang}}, \ and\ \bibinfo {author} {\bibfnamefont {Andreas}\
  \bibnamefont {Albrecht}},\ }\bibfield  {title} {\enquote {\bibinfo {title}
  {Critical open-system dynamics in a one-dimensional optical-lattice clock},}\
  }\href {\doibase 10.1103/PhysRevA.99.023802} {\bibfield  {journal} {\bibinfo
  {journal} {Phys. Rev. A}\ }\textbf {\bibinfo {volume} {99}},\ \bibinfo
  {pages} {023802} (\bibinfo {year} {2019})}\BibitemShut {NoStop}%
\bibitem [{\citenamefont {Plankensteiner}\ \emph {et~al.}(2019)\citenamefont
  {Plankensteiner}, \citenamefont {Sommer}, \citenamefont {Reitz},
  \citenamefont {Ritsch},\ and\ \citenamefont {Genes}}]{Plankensteiner19}%
  \BibitemOpen
  \bibfield  {author} {\bibinfo {author} {\bibfnamefont {D.}~\bibnamefont
  {Plankensteiner}}, \bibinfo {author} {\bibfnamefont {C.}~\bibnamefont
  {Sommer}}, \bibinfo {author} {\bibfnamefont {M.}~\bibnamefont {Reitz}},
  \bibinfo {author} {\bibfnamefont {H.}~\bibnamefont {Ritsch}}, \ and\ \bibinfo
  {author} {\bibfnamefont {C.}~\bibnamefont {Genes}},\ }\bibfield  {title}
  {\enquote {\bibinfo {title} {Enhanced collective purcell effect of coupled
  quantum emitter systems},}\ }\href {\doibase 10.1103/PhysRevA.99.043843}
  {\bibfield  {journal} {\bibinfo  {journal} {Phys. Rev. A}\ }\textbf {\bibinfo
  {volume} {99}},\ \bibinfo {pages} {043843} (\bibinfo {year}
  {2019})}\BibitemShut {NoStop}%
\bibitem [{\citenamefont {Javanainen}\ and\ \citenamefont
  {Rajapakse}(2019)}]{Javanainen19}%
  \BibitemOpen
  \bibfield  {author} {\bibinfo {author} {\bibfnamefont {Juha}\ \bibnamefont
  {Javanainen}}\ and\ \bibinfo {author} {\bibfnamefont {Renuka}\ \bibnamefont
  {Rajapakse}},\ }\bibfield  {title} {\enquote {\bibinfo {title} {Light
  propagation in systems involving two-dimensional atomic lattices},}\ }\href
  {\doibase 10.1103/PhysRevA.100.013616} {\bibfield  {journal} {\bibinfo
  {journal} {Phys. Rev. A}\ }\textbf {\bibinfo {volume} {100}},\ \bibinfo
  {pages} {013616} (\bibinfo {year} {2019})}\BibitemShut {NoStop}%
\bibitem [{\citenamefont {Bettles}\ \emph {et~al.}(2020)\citenamefont
  {Bettles}, \citenamefont {Lee}, \citenamefont {Gardiner},\ and\ \citenamefont
  {Ruostekoski}}]{Bettles20}%
  \BibitemOpen
  \bibfield  {author} {\bibinfo {author} {\bibfnamefont {Robert~J.}\
  \bibnamefont {Bettles}}, \bibinfo {author} {\bibfnamefont {Mark~D.}\
  \bibnamefont {Lee}}, \bibinfo {author} {\bibfnamefont {Simon~A.}\
  \bibnamefont {Gardiner}}, \ and\ \bibinfo {author} {\bibfnamefont {Janne}\
  \bibnamefont {Ruostekoski}},\ }\bibfield  {title} {\enquote {\bibinfo {title}
  {Quantum and nonlinear effects in light transmitted through planar atomic
  arrays},}\ }\href {\doibase 10.1038/s42005-020-00404-3} {\bibfield  {journal}
  {\bibinfo  {journal} {Communications Physics}\ }\textbf {\bibinfo {volume}
  {3}},\ \bibinfo {pages} {141} (\bibinfo {year} {2020})}\BibitemShut {NoStop}%
\bibitem [{\citenamefont {Qu}\ and\ \citenamefont {Rey}(2019)}]{Qu19}%
  \BibitemOpen
  \bibfield  {author} {\bibinfo {author} {\bibfnamefont {Chunlei}\ \bibnamefont
  {Qu}}\ and\ \bibinfo {author} {\bibfnamefont {Ana~M.}\ \bibnamefont {Rey}},\
  }\bibfield  {title} {\enquote {\bibinfo {title} {Spin squeezing and many-body
  dipolar dynamics in optical lattice clocks},}\ }\href {\doibase
  10.1103/PhysRevA.100.041602} {\bibfield  {journal} {\bibinfo  {journal}
  {Phys. Rev. A}\ }\textbf {\bibinfo {volume} {100}},\ \bibinfo {pages}
  {041602(R)} (\bibinfo {year} {2019})}\BibitemShut {NoStop}%
\bibitem [{\citenamefont {Zhang}\ \emph {et~al.}(2020)\citenamefont {Zhang},
  \citenamefont {Yu},\ and\ \citenamefont {M\o{}lmer}}]{Zhang20}%
  \BibitemOpen
  \bibfield  {author} {\bibinfo {author} {\bibfnamefont {Yu-Xiang}\
  \bibnamefont {Zhang}}, \bibinfo {author} {\bibfnamefont {Chuan}\ \bibnamefont
  {Yu}}, \ and\ \bibinfo {author} {\bibfnamefont {Klaus}\ \bibnamefont
  {M\o{}lmer}},\ }\bibfield  {title} {\enquote {\bibinfo {title} {Subradiant
  bound dimer excited states of emitter chains coupled to a one dimensional
  waveguide},}\ }\href {\doibase 10.1103/PhysRevResearch.2.013173} {\bibfield
  {journal} {\bibinfo  {journal} {Phys. Rev. Research}\ }\textbf {\bibinfo
  {volume} {2}},\ \bibinfo {pages} {013173} (\bibinfo {year}
  {2020})}\BibitemShut {NoStop}%
\bibitem [{\citenamefont {Lemoult}\ \emph {et~al.}(2010)\citenamefont
  {Lemoult}, \citenamefont {Lerosey}, \citenamefont {de~Rosny},\ and\
  \citenamefont {Fink}}]{lemoult2010}%
  \BibitemOpen
  \bibfield  {author} {\bibinfo {author} {\bibfnamefont {Fabrice}\ \bibnamefont
  {Lemoult}}, \bibinfo {author} {\bibfnamefont {Geoffroy}\ \bibnamefont
  {Lerosey}}, \bibinfo {author} {\bibfnamefont {Julien}\ \bibnamefont
  {de~Rosny}}, \ and\ \bibinfo {author} {\bibfnamefont {Mathias}\ \bibnamefont
  {Fink}},\ }\bibfield  {title} {\enquote {\bibinfo {title} {Resonant
  metalenses for breaking the diffraction barrier},}\ }\href {\doibase
  10.1103/PhysRevLett.104.203901} {\bibfield  {journal} {\bibinfo  {journal}
  {Phys. Rev. Lett.}\ }\textbf {\bibinfo {volume} {104}},\ \bibinfo {pages}
  {203901} (\bibinfo {year} {2010})}\BibitemShut {NoStop}%
\bibitem [{\citenamefont {Jenkins}\ \emph {et~al.}(2017)\citenamefont
  {Jenkins}, \citenamefont {Ruostekoski}, \citenamefont {Papasimakis},
  \citenamefont {Savo},\ and\ \citenamefont {Zheludev}}]{Jenkins17}%
  \BibitemOpen
  \bibfield  {author} {\bibinfo {author} {\bibfnamefont {Stewart~D.}\
  \bibnamefont {Jenkins}}, \bibinfo {author} {\bibfnamefont {Janne}\
  \bibnamefont {Ruostekoski}}, \bibinfo {author} {\bibfnamefont {Nikitas}\
  \bibnamefont {Papasimakis}}, \bibinfo {author} {\bibfnamefont {Salvatore}\
  \bibnamefont {Savo}}, \ and\ \bibinfo {author} {\bibfnamefont {Nikolay~I.}\
  \bibnamefont {Zheludev}},\ }\bibfield  {title} {\enquote {\bibinfo {title}
  {Many-body subradiant excitations in metamaterial arrays: Experiment and
  theory},}\ }\href {\doibase 10.1103/PhysRevLett.119.053901} {\bibfield
  {journal} {\bibinfo  {journal} {Phys. Rev. Lett.}\ }\textbf {\bibinfo
  {volume} {119}},\ \bibinfo {pages} {053901} (\bibinfo {year}
  {2017})}\BibitemShut {NoStop}%
\bibitem [{\citenamefont {Grankin}\ \emph {et~al.}(2018)\citenamefont
  {Grankin}, \citenamefont {Guimond}, \citenamefont {Vasilyev}, \citenamefont
  {Vermersch},\ and\ \citenamefont {Zoller}}]{Grankin18}%
  \BibitemOpen
  \bibfield  {author} {\bibinfo {author} {\bibfnamefont {A.}~\bibnamefont
  {Grankin}}, \bibinfo {author} {\bibfnamefont {P.~O.}\ \bibnamefont
  {Guimond}}, \bibinfo {author} {\bibfnamefont {D.~V.}\ \bibnamefont
  {Vasilyev}}, \bibinfo {author} {\bibfnamefont {B.}~\bibnamefont {Vermersch}},
  \ and\ \bibinfo {author} {\bibfnamefont {P.}~\bibnamefont {Zoller}},\
  }\bibfield  {title} {\enquote {\bibinfo {title} {Free-space photonic quantum
  link and chiral quantum optics},}\ }\href {\doibase
  10.1103/PhysRevA.98.043825} {\bibfield  {journal} {\bibinfo  {journal} {Phys.
  Rev. A}\ }\textbf {\bibinfo {volume} {98}},\ \bibinfo {pages} {043825}
  (\bibinfo {year} {2018})}\BibitemShut {NoStop}%
\bibitem [{\citenamefont {Ballantine}\ and\ \citenamefont
  {Ruostekoski}(2020)}]{Ballantine20ant}%
  \BibitemOpen
  \bibfield  {author} {\bibinfo {author} {\bibfnamefont {K.~E.}\ \bibnamefont
  {Ballantine}}\ and\ \bibinfo {author} {\bibfnamefont {J.}~\bibnamefont
  {Ruostekoski}},\ }\bibfield  {title} {\enquote {\bibinfo {title}
  {Subradiance-protected excitation spreading in the generation of collimated
  photon emission from an atomic array},}\ }\href {\doibase
  10.1103/PhysRevResearch.2.023086} {\bibfield  {journal} {\bibinfo  {journal}
  {Phys. Rev. Research}\ }\textbf {\bibinfo {volume} {2}},\ \bibinfo {pages}
  {023086} (\bibinfo {year} {2020})}\BibitemShut {NoStop}%
\bibitem [{\citenamefont {Balik}\ \emph {et~al.}(2013)\citenamefont {Balik},
  \citenamefont {Win}, \citenamefont {Havey}, \citenamefont {Sokolov},\ and\
  \citenamefont {Kupriyanov}}]{BalikEtAl2013}%
  \BibitemOpen
  \bibfield  {author} {\bibinfo {author} {\bibfnamefont {S.}~\bibnamefont
  {Balik}}, \bibinfo {author} {\bibfnamefont {A.~L.}\ \bibnamefont {Win}},
  \bibinfo {author} {\bibfnamefont {M.~D.}\ \bibnamefont {Havey}}, \bibinfo
  {author} {\bibfnamefont {I.~M.}\ \bibnamefont {Sokolov}}, \ and\ \bibinfo
  {author} {\bibfnamefont {D.~V.}\ \bibnamefont {Kupriyanov}},\ }\bibfield
  {title} {\enquote {\bibinfo {title} {Near-resonance light scattering from a
  high-density ultracold atomic ${}^{87}${Rb} gas},}\ }\href {\doibase
  10.1103/PhysRevA.87.053817} {\bibfield  {journal} {\bibinfo  {journal} {Phys.
  Rev. A}\ }\textbf {\bibinfo {volume} {87}},\ \bibinfo {pages} {053817}
  (\bibinfo {year} {2013})}\BibitemShut {NoStop}%
\bibitem [{\citenamefont {Chab\'e}\ \emph {et~al.}(2014)\citenamefont
  {Chab\'e}, \citenamefont {Rouabah}, \citenamefont {Bellando}, \citenamefont
  {Bienaim\'e}, \citenamefont {Piovella}, \citenamefont {Bachelard},\ and\
  \citenamefont {Kaiser}}]{CHA14}%
  \BibitemOpen
  \bibfield  {author} {\bibinfo {author} {\bibfnamefont {Julien}\ \bibnamefont
  {Chab\'e}}, \bibinfo {author} {\bibfnamefont {Mohamed-Taha}\ \bibnamefont
  {Rouabah}}, \bibinfo {author} {\bibfnamefont {Louis}\ \bibnamefont
  {Bellando}}, \bibinfo {author} {\bibfnamefont {Tom}\ \bibnamefont
  {Bienaim\'e}}, \bibinfo {author} {\bibfnamefont {Nicola}\ \bibnamefont
  {Piovella}}, \bibinfo {author} {\bibfnamefont {Romain}\ \bibnamefont
  {Bachelard}}, \ and\ \bibinfo {author} {\bibfnamefont {Robin}\ \bibnamefont
  {Kaiser}},\ }\bibfield  {title} {\enquote {\bibinfo {title} {Coherent and
  incoherent multiple scattering},}\ }\href {\doibase
  10.1103/PhysRevA.89.043833} {\bibfield  {journal} {\bibinfo  {journal} {Phys.
  Rev. A}\ }\textbf {\bibinfo {volume} {89}},\ \bibinfo {pages} {043833}
  (\bibinfo {year} {2014})}\BibitemShut {NoStop}%
\bibitem [{\citenamefont {Pellegrino}\ \emph {et~al.}(2014)\citenamefont
  {Pellegrino}, \citenamefont {Bourgain}, \citenamefont {Jennewein},
  \citenamefont {Sortais}, \citenamefont {Browaeys}, \citenamefont {Jenkins},\
  and\ \citenamefont {Ruostekoski}}]{Pellegrino2014a}%
  \BibitemOpen
  \bibfield  {author} {\bibinfo {author} {\bibfnamefont {J.}~\bibnamefont
  {Pellegrino}}, \bibinfo {author} {\bibfnamefont {R.}~\bibnamefont
  {Bourgain}}, \bibinfo {author} {\bibfnamefont {S.}~\bibnamefont {Jennewein}},
  \bibinfo {author} {\bibfnamefont {Y.~R.~P.}\ \bibnamefont {Sortais}},
  \bibinfo {author} {\bibfnamefont {A.}~\bibnamefont {Browaeys}}, \bibinfo
  {author} {\bibfnamefont {S.~D.}\ \bibnamefont {Jenkins}}, \ and\ \bibinfo
  {author} {\bibfnamefont {J.}~\bibnamefont {Ruostekoski}},\ }\bibfield
  {title} {\enquote {\bibinfo {title} {Observation of suppression of light
  scattering induced by dipole-dipole interactions in a cold-atom ensemble},}\
  }\href {\doibase 10.1103/PhysRevLett.113.133602} {\bibfield  {journal}
  {\bibinfo  {journal} {Phys. Rev. Lett.}\ }\textbf {\bibinfo {volume} {113}},\
  \bibinfo {pages} {133602} (\bibinfo {year} {2014})}\BibitemShut {NoStop}%
\bibitem [{\citenamefont {Sheremet}\ \emph {et~al.}(2014)\citenamefont
  {Sheremet}, \citenamefont {Sokolov}, \citenamefont {Kupriyanov},
  \citenamefont {Balik}, \citenamefont {Win},\ and\ \citenamefont
  {Havey}}]{Havey_jmo14}%
  \BibitemOpen
  \bibfield  {author} {\bibinfo {author} {\bibfnamefont {A.S.}\ \bibnamefont
  {Sheremet}}, \bibinfo {author} {\bibfnamefont {I.M.}\ \bibnamefont
  {Sokolov}}, \bibinfo {author} {\bibfnamefont {D.V.}\ \bibnamefont
  {Kupriyanov}}, \bibinfo {author} {\bibfnamefont {S.}~\bibnamefont {Balik}},
  \bibinfo {author} {\bibfnamefont {A.L.}\ \bibnamefont {Win}}, \ and\ \bibinfo
  {author} {\bibfnamefont {M.D.}\ \bibnamefont {Havey}},\ }\bibfield  {title}
  {\enquote {\bibinfo {title} {Light scattering on the {$F = 1 \rightarrow F' =
  0$} transition in a cold and high density {$^{87}$Rb} vapor},}\ }\href
  {\doibase 10.1080/09500340.2013.854419} {\bibfield  {journal} {\bibinfo
  {journal} {Journal of Modern Optics}\ }\textbf {\bibinfo {volume} {61}},\
  \bibinfo {pages} {77--84} (\bibinfo {year} {2014})}\BibitemShut {NoStop}%
\bibitem [{\citenamefont {Kwong}\ \emph {et~al.}(2014)\citenamefont {Kwong},
  \citenamefont {Yang}, \citenamefont {Pramod}, \citenamefont {Pandey},
  \citenamefont {Delande}, \citenamefont {Pierrat},\ and\ \citenamefont
  {Wilkowski}}]{wilkowski}%
  \BibitemOpen
  \bibfield  {author} {\bibinfo {author} {\bibfnamefont {C.~C.}\ \bibnamefont
  {Kwong}}, \bibinfo {author} {\bibfnamefont {T.}~\bibnamefont {Yang}},
  \bibinfo {author} {\bibfnamefont {M.~S.}\ \bibnamefont {Pramod}}, \bibinfo
  {author} {\bibfnamefont {K.}~\bibnamefont {Pandey}}, \bibinfo {author}
  {\bibfnamefont {D.}~\bibnamefont {Delande}}, \bibinfo {author} {\bibfnamefont
  {R.}~\bibnamefont {Pierrat}}, \ and\ \bibinfo {author} {\bibfnamefont
  {D.}~\bibnamefont {Wilkowski}},\ }\bibfield  {title} {\enquote {\bibinfo
  {title} {Cooperative emission of a coherent superflash of light},}\ }\href
  {\doibase 10.1103/PhysRevLett.113.223601} {\bibfield  {journal} {\bibinfo
  {journal} {Phys. Rev. Lett.}\ }\textbf {\bibinfo {volume} {113}},\ \bibinfo
  {pages} {223601} (\bibinfo {year} {2014})}\BibitemShut {NoStop}%
\bibitem [{\citenamefont {Jennewein}\ \emph {et~al.}(2016)\citenamefont
  {Jennewein}, \citenamefont {Besbes}, \citenamefont {Schilder}, \citenamefont
  {Jenkins}, \citenamefont {Sauvan}, \citenamefont {Ruostekoski}, \citenamefont
  {Greffet}, \citenamefont {Sortais},\ and\ \citenamefont
  {Browaeys}}]{Jennewein_trans}%
  \BibitemOpen
  \bibfield  {author} {\bibinfo {author} {\bibfnamefont {S.}~\bibnamefont
  {Jennewein}}, \bibinfo {author} {\bibfnamefont {M.}~\bibnamefont {Besbes}},
  \bibinfo {author} {\bibfnamefont {N.~J.}\ \bibnamefont {Schilder}}, \bibinfo
  {author} {\bibfnamefont {S.~D.}\ \bibnamefont {Jenkins}}, \bibinfo {author}
  {\bibfnamefont {C.}~\bibnamefont {Sauvan}}, \bibinfo {author} {\bibfnamefont
  {J.}~\bibnamefont {Ruostekoski}}, \bibinfo {author} {\bibfnamefont {J.-J.}\
  \bibnamefont {Greffet}}, \bibinfo {author} {\bibfnamefont {Y.~R.~P.}\
  \bibnamefont {Sortais}}, \ and\ \bibinfo {author} {\bibfnamefont
  {A.}~\bibnamefont {Browaeys}},\ }\bibfield  {title} {\enquote {\bibinfo
  {title} {Coherent scattering of near-resonant light by a dense microscopic
  cold atomic cloud},}\ }\href {\doibase 10.1103/PhysRevLett.116.233601}
  {\bibfield  {journal} {\bibinfo  {journal} {Phys. Rev. Lett.}\ }\textbf
  {\bibinfo {volume} {116}},\ \bibinfo {pages} {233601} (\bibinfo {year}
  {2016})}\BibitemShut {NoStop}%
\bibitem [{\citenamefont {Bromley}\ \emph {et~al.}(2016)\citenamefont
  {Bromley}, \citenamefont {Zhu}, \citenamefont {Bishof}, \citenamefont
  {Zhang}, \citenamefont {Bothwell}, \citenamefont {Schachenmayer},
  \citenamefont {Nicholson}, \citenamefont {Kaiser}, \citenamefont {Yelin},
  \citenamefont {Lukin}, \citenamefont {Rey},\ and\ \citenamefont
  {Ye}}]{Ye2016}%
  \BibitemOpen
  \bibfield  {author} {\bibinfo {author} {\bibfnamefont {S.~L.}\ \bibnamefont
  {Bromley}}, \bibinfo {author} {\bibfnamefont {B.}~\bibnamefont {Zhu}},
  \bibinfo {author} {\bibfnamefont {M.}~\bibnamefont {Bishof}}, \bibinfo
  {author} {\bibfnamefont {X.}~\bibnamefont {Zhang}}, \bibinfo {author}
  {\bibfnamefont {T.}~\bibnamefont {Bothwell}}, \bibinfo {author}
  {\bibfnamefont {J.}~\bibnamefont {Schachenmayer}}, \bibinfo {author}
  {\bibfnamefont {T.~L.}\ \bibnamefont {Nicholson}}, \bibinfo {author}
  {\bibfnamefont {R.}~\bibnamefont {Kaiser}}, \bibinfo {author} {\bibfnamefont
  {S.~F.}\ \bibnamefont {Yelin}}, \bibinfo {author} {\bibfnamefont {M.~D.}\
  \bibnamefont {Lukin}}, \bibinfo {author} {\bibfnamefont {A.~M.}\ \bibnamefont
  {Rey}}, \ and\ \bibinfo {author} {\bibfnamefont {J.}~\bibnamefont {Ye}},\
  }\bibfield  {title} {\enquote {\bibinfo {title} {Collective atomic scattering
  and motional effects in a dense coherent medium},}\ }\href
  {http://dx.doi.org/10.1038/ncomms11039} {\bibfield  {journal} {\bibinfo
  {journal} {Nat Commun}\ }\textbf {\bibinfo {volume} {7}},\ \bibinfo {pages}
  {11039} (\bibinfo {year} {2016})}\BibitemShut {NoStop}%
\bibitem [{\citenamefont {Jenkins}\ \emph
  {et~al.}(2016{\natexlab{a}})\citenamefont {Jenkins}, \citenamefont
  {Ruostekoski}, \citenamefont {Javanainen}, \citenamefont {Bourgain},
  \citenamefont {Jennewein}, \citenamefont {Sortais},\ and\ \citenamefont
  {Browaeys}}]{Jenkins_thermshift}%
  \BibitemOpen
  \bibfield  {author} {\bibinfo {author} {\bibfnamefont {S.~D.}\ \bibnamefont
  {Jenkins}}, \bibinfo {author} {\bibfnamefont {J.}~\bibnamefont
  {Ruostekoski}}, \bibinfo {author} {\bibfnamefont {J.}~\bibnamefont
  {Javanainen}}, \bibinfo {author} {\bibfnamefont {R.}~\bibnamefont
  {Bourgain}}, \bibinfo {author} {\bibfnamefont {S.}~\bibnamefont {Jennewein}},
  \bibinfo {author} {\bibfnamefont {Y.~R.~P.}\ \bibnamefont {Sortais}}, \ and\
  \bibinfo {author} {\bibfnamefont {A.}~\bibnamefont {Browaeys}},\ }\bibfield
  {title} {\enquote {\bibinfo {title} {Optical resonance shifts in the
  fluorescence of thermal and cold atomic gases},}\ }\href {\doibase
  10.1103/PhysRevLett.116.183601} {\bibfield  {journal} {\bibinfo  {journal}
  {Phys. Rev. Lett.}\ }\textbf {\bibinfo {volume} {116}},\ \bibinfo {pages}
  {183601} (\bibinfo {year} {2016}{\natexlab{a}})}\BibitemShut {NoStop}%
\bibitem [{\citenamefont {Bons}\ \emph {et~al.}(2016)\citenamefont {Bons},
  \citenamefont {de~Haas}, \citenamefont {de~Jong}, \citenamefont {Groot},\
  and\ \citenamefont {van~der Straten}}]{vdStraten16}%
  \BibitemOpen
  \bibfield  {author} {\bibinfo {author} {\bibfnamefont {P.~C.}\ \bibnamefont
  {Bons}}, \bibinfo {author} {\bibfnamefont {R.}~\bibnamefont {de~Haas}},
  \bibinfo {author} {\bibfnamefont {D.}~\bibnamefont {de~Jong}}, \bibinfo
  {author} {\bibfnamefont {A.}~\bibnamefont {Groot}}, \ and\ \bibinfo {author}
  {\bibfnamefont {P.}~\bibnamefont {van~der Straten}},\ }\bibfield  {title}
  {\enquote {\bibinfo {title} {Quantum enhancement of the index of refraction
  in a {Bose-Einstein} condensate},}\ }\href {\doibase
  10.1103/PhysRevLett.116.173602} {\bibfield  {journal} {\bibinfo  {journal}
  {Phys. Rev. Lett.}\ }\textbf {\bibinfo {volume} {116}},\ \bibinfo {pages}
  {173602} (\bibinfo {year} {2016})}\BibitemShut {NoStop}%
\bibitem [{\citenamefont {Guerin}\ \emph {et~al.}(2016)\citenamefont {Guerin},
  \citenamefont {Ara\'ujo},\ and\ \citenamefont {Kaiser}}]{Guerin_subr16}%
  \BibitemOpen
  \bibfield  {author} {\bibinfo {author} {\bibfnamefont {William}\ \bibnamefont
  {Guerin}}, \bibinfo {author} {\bibfnamefont {Michelle~O.}\ \bibnamefont
  {Ara\'ujo}}, \ and\ \bibinfo {author} {\bibfnamefont {Robin}\ \bibnamefont
  {Kaiser}},\ }\bibfield  {title} {\enquote {\bibinfo {title} {Subradiance in a
  large cloud of cold atoms},}\ }\href {\doibase
  10.1103/PhysRevLett.116.083601} {\bibfield  {journal} {\bibinfo  {journal}
  {Phys. Rev. Lett.}\ }\textbf {\bibinfo {volume} {116}},\ \bibinfo {pages}
  {083601} (\bibinfo {year} {2016})}\BibitemShut {NoStop}%
\bibitem [{\citenamefont {Machluf}\ \emph {et~al.}(2019)\citenamefont
  {Machluf}, \citenamefont {Naber}, \citenamefont {Soudijn}, \citenamefont
  {Ruostekoski},\ and\ \citenamefont {Spreeuw}}]{Machluf2018}%
  \BibitemOpen
  \bibfield  {author} {\bibinfo {author} {\bibfnamefont {Shimon}\ \bibnamefont
  {Machluf}}, \bibinfo {author} {\bibfnamefont {Julian~B.}\ \bibnamefont
  {Naber}}, \bibinfo {author} {\bibfnamefont {Maarten~L.}\ \bibnamefont
  {Soudijn}}, \bibinfo {author} {\bibfnamefont {Janne}\ \bibnamefont
  {Ruostekoski}}, \ and\ \bibinfo {author} {\bibfnamefont {Robert J.~C.}\
  \bibnamefont {Spreeuw}},\ }\bibfield  {title} {\enquote {\bibinfo {title}
  {Collective suppression of optical hyperfine pumping in dense clouds of atoms
  in microtraps},}\ }\href {\doibase 10.1103/PhysRevA.100.051801} {\bibfield
  {journal} {\bibinfo  {journal} {Phys. Rev. A}\ }\textbf {\bibinfo {volume}
  {100}},\ \bibinfo {pages} {051801(R)} (\bibinfo {year} {2019})}\BibitemShut
  {NoStop}%
\bibitem [{\citenamefont {Corman}\ \emph {et~al.}(2017)\citenamefont {Corman},
  \citenamefont {Ville}, \citenamefont {Saint-Jalm}, \citenamefont
  {Aidelsburger}, \citenamefont {Bienaim\'e}, \citenamefont {Nascimb\`ene},
  \citenamefont {Dalibard},\ and\ \citenamefont {Beugnon}}]{Dalibard_slab}%
  \BibitemOpen
  \bibfield  {author} {\bibinfo {author} {\bibfnamefont {L.}~\bibnamefont
  {Corman}}, \bibinfo {author} {\bibfnamefont {J.~L.}\ \bibnamefont {Ville}},
  \bibinfo {author} {\bibfnamefont {R.}~\bibnamefont {Saint-Jalm}}, \bibinfo
  {author} {\bibfnamefont {M.}~\bibnamefont {Aidelsburger}}, \bibinfo {author}
  {\bibfnamefont {T.}~\bibnamefont {Bienaim\'e}}, \bibinfo {author}
  {\bibfnamefont {S.}~\bibnamefont {Nascimb\`ene}}, \bibinfo {author}
  {\bibfnamefont {J.}~\bibnamefont {Dalibard}}, \ and\ \bibinfo {author}
  {\bibfnamefont {J.}~\bibnamefont {Beugnon}},\ }\bibfield  {title} {\enquote
  {\bibinfo {title} {Transmission of near-resonant light through a dense slab
  of cold atoms},}\ }\href {\doibase 10.1103/PhysRevA.96.053629} {\bibfield
  {journal} {\bibinfo  {journal} {Phys. Rev. A}\ }\textbf {\bibinfo {volume}
  {96}},\ \bibinfo {pages} {053629} (\bibinfo {year} {2017})}\BibitemShut
  {NoStop}%
\bibitem [{\citenamefont {Bettles}\ \emph {et~al.}(2018)\citenamefont
  {Bettles}, \citenamefont {Ilieva}, \citenamefont {Busche}, \citenamefont
  {Huillery}, \citenamefont {Ball}, \citenamefont {Spong},\ and\ \citenamefont
  {Adams}}]{Bettles18}%
  \BibitemOpen
  \bibfield  {author} {\bibinfo {author} {\bibfnamefont {R.~J.}\ \bibnamefont
  {Bettles}}, \bibinfo {author} {\bibfnamefont {T.}~\bibnamefont {Ilieva}},
  \bibinfo {author} {\bibfnamefont {H.}~\bibnamefont {Busche}}, \bibinfo
  {author} {\bibfnamefont {P.}~\bibnamefont {Huillery}}, \bibinfo {author}
  {\bibfnamefont {S.~W.}\ \bibnamefont {Ball}}, \bibinfo {author}
  {\bibfnamefont {N.~L.~R.}\ \bibnamefont {Spong}}, \ and\ \bibinfo {author}
  {\bibfnamefont {C.~S.}\ \bibnamefont {Adams}},\ }\bibfield  {title} {\enquote
  {\bibinfo {title} {Collective mode interferences in light-matter
  interactions},}\ }\href@noop {} {\  (\bibinfo {year} {2018})},\ \Eprint
  {http://arxiv.org/abs/arXiv:1808.08415} {arXiv:1808.08415} \BibitemShut
  {NoStop}%
\bibitem [{\citenamefont {Rui}\ \emph {et~al.}(2020)\citenamefont {Rui},
  \citenamefont {Wei}, \citenamefont {Rubio-Abadal}, \citenamefont {Hollerith},
  \citenamefont {Zeiher}, \citenamefont {Stamper-Kurn}, \citenamefont {Gross},\
  and\ \citenamefont {Bloch}}]{rui2020}%
  \BibitemOpen
  \bibfield  {author} {\bibinfo {author} {\bibfnamefont {Jun}\ \bibnamefont
  {Rui}}, \bibinfo {author} {\bibfnamefont {David}\ \bibnamefont {Wei}},
  \bibinfo {author} {\bibfnamefont {Antonio}\ \bibnamefont {Rubio-Abadal}},
  \bibinfo {author} {\bibfnamefont {Simon}\ \bibnamefont {Hollerith}}, \bibinfo
  {author} {\bibfnamefont {Johannes}\ \bibnamefont {Zeiher}}, \bibinfo {author}
  {\bibfnamefont {Dan~M.}\ \bibnamefont {Stamper-Kurn}}, \bibinfo {author}
  {\bibfnamefont {Christian}\ \bibnamefont {Gross}}, \ and\ \bibinfo {author}
  {\bibfnamefont {Immanuel}\ \bibnamefont {Bloch}},\ }\bibfield  {title}
  {\enquote {\bibinfo {title} {{A subradiant optical mirror formed by a single
  structured atomic layer}},}\ }\href {\doibase 10.1038/s41586-020-2463-x}
  {\bibfield  {journal} {\bibinfo  {journal} {Nature}\ }\textbf {\bibinfo
  {volume} {583}},\ \bibinfo {pages} {369--374} (\bibinfo {year}
  {2020})}\BibitemShut {NoStop}%
\bibitem [{\citenamefont {Mair}\ \emph {et~al.}(2001)\citenamefont {Mair},
  \citenamefont {Vaziri}, \citenamefont {Weihs},\ and\ \citenamefont
  {Zeilinger}}]{Mair01}%
  \BibitemOpen
  \bibfield  {author} {\bibinfo {author} {\bibfnamefont {Alois}\ \bibnamefont
  {Mair}}, \bibinfo {author} {\bibfnamefont {Alipasha}\ \bibnamefont {Vaziri}},
  \bibinfo {author} {\bibfnamefont {Gregor}\ \bibnamefont {Weihs}}, \ and\
  \bibinfo {author} {\bibfnamefont {Anton}\ \bibnamefont {Zeilinger}},\
  }\bibfield  {title} {\enquote {\bibinfo {title} {Entanglement of the orbital
  angular momentum states of photons},}\ }\href {\doibase 10.1038/35085529}
  {\bibfield  {journal} {\bibinfo  {journal} {Nature}\ }\textbf {\bibinfo
  {volume} {412}},\ \bibinfo {pages} {313--316} (\bibinfo {year}
  {2001})}\BibitemShut {NoStop}%
\bibitem [{\citenamefont {Bechmann-Pasquinucci}\ and\ \citenamefont
  {Tittel}(2000)}]{Bechmann00}%
  \BibitemOpen
  \bibfield  {author} {\bibinfo {author} {\bibfnamefont {H.}~\bibnamefont
  {Bechmann-Pasquinucci}}\ and\ \bibinfo {author} {\bibfnamefont
  {W.}~\bibnamefont {Tittel}},\ }\bibfield  {title} {\enquote {\bibinfo {title}
  {Quantum cryptography using larger alphabets},}\ }\href {\doibase
  10.1103/PhysRevA.61.062308} {\bibfield  {journal} {\bibinfo  {journal} {Phys.
  Rev. A}\ }\textbf {\bibinfo {volume} {61}},\ \bibinfo {pages} {062308}
  (\bibinfo {year} {2000})}\BibitemShut {NoStop}%
\bibitem [{\citenamefont {Pfeiffer}\ and\ \citenamefont
  {Grbic}(2013)}]{Pfeiffer13}%
  \BibitemOpen
  \bibfield  {author} {\bibinfo {author} {\bibfnamefont {Carl}\ \bibnamefont
  {Pfeiffer}}\ and\ \bibinfo {author} {\bibfnamefont {Anthony}\ \bibnamefont
  {Grbic}},\ }\bibfield  {title} {\enquote {\bibinfo {title} {Metamaterial
  huygens' surfaces: Tailoring wave fronts with reflectionless sheets},}\
  }\href {\doibase 10.1103/PhysRevLett.110.197401} {\bibfield  {journal}
  {\bibinfo  {journal} {Phys. Rev. Lett.}\ }\textbf {\bibinfo {volume} {110}},\
  \bibinfo {pages} {197401} (\bibinfo {year} {2013})}\BibitemShut {NoStop}%
\bibitem [{\citenamefont {Decker}\ \emph {et~al.}(2015)\citenamefont {Decker},
  \citenamefont {Staude}, \citenamefont {Falkner}, \citenamefont {Dominguez},
  \citenamefont {Neshev}, \citenamefont {Brener}, \citenamefont {Pertsch},\
  and\ \citenamefont {Kivshar}}]{Decker15}%
  \BibitemOpen
  \bibfield  {author} {\bibinfo {author} {\bibfnamefont {Manuel}\ \bibnamefont
  {Decker}}, \bibinfo {author} {\bibfnamefont {Isabelle}\ \bibnamefont
  {Staude}}, \bibinfo {author} {\bibfnamefont {Matthias}\ \bibnamefont
  {Falkner}}, \bibinfo {author} {\bibfnamefont {Jason}\ \bibnamefont
  {Dominguez}}, \bibinfo {author} {\bibfnamefont {Dragomir~N.}\ \bibnamefont
  {Neshev}}, \bibinfo {author} {\bibfnamefont {Igal}\ \bibnamefont {Brener}},
  \bibinfo {author} {\bibfnamefont {Thomas}\ \bibnamefont {Pertsch}}, \ and\
  \bibinfo {author} {\bibfnamefont {Yuri~S.}\ \bibnamefont {Kivshar}},\
  }\bibfield  {title} {\enquote {\bibinfo {title} {High-efficiency dielectric
  huygens’ surfaces},}\ }\href {\doibase 10.1002/adom.201400584} {\bibfield
  {journal} {\bibinfo  {journal} {Advanced Optical Materials}\ }\textbf
  {\bibinfo {volume} {3}},\ \bibinfo {pages} {813--820} (\bibinfo {year}
  {2015})}\BibitemShut {NoStop}%
\bibitem [{\citenamefont {Yu}\ \emph {et~al.}(2015)\citenamefont {Yu},
  \citenamefont {Zhu}, \citenamefont {Paniagua-Domínguez}, \citenamefont {Fu},
  \citenamefont {Luk'yanchuk},\ and\ \citenamefont {Kuznetsov}}]{Yu15}%
  \BibitemOpen
  \bibfield  {author} {\bibinfo {author} {\bibfnamefont {Ye~Feng}\ \bibnamefont
  {Yu}}, \bibinfo {author} {\bibfnamefont {Alexander~Y.}\ \bibnamefont {Zhu}},
  \bibinfo {author} {\bibfnamefont {Ramón}\ \bibnamefont
  {Paniagua-Domínguez}}, \bibinfo {author} {\bibfnamefont {Yuan~Hsing}\
  \bibnamefont {Fu}}, \bibinfo {author} {\bibfnamefont {Boris}\ \bibnamefont
  {Luk'yanchuk}}, \ and\ \bibinfo {author} {\bibfnamefont {Arseniy~I.}\
  \bibnamefont {Kuznetsov}},\ }\bibfield  {title} {\enquote {\bibinfo {title}
  {High-transmission dielectric metasurface with {$2\pi$} phase control at
  visible wavelengths},}\ }\href
  {https://onlinelibrary.wiley.com/doi/abs/10.1002/lpor.201500041} {\bibfield
  {journal} {\bibinfo  {journal} {Laser \& Photonics Reviews}\ }\textbf
  {\bibinfo {volume} {9}},\ \bibinfo {pages} {412--418} (\bibinfo {year}
  {2015})}\BibitemShut {NoStop}%
\bibitem [{\citenamefont {Chong}\ \emph {et~al.}(2015)\citenamefont {Chong},
  \citenamefont {Staude}, \citenamefont {James}, \citenamefont {Dominguez},
  \citenamefont {Liu}, \citenamefont {Campione}, \citenamefont {Subramania},
  \citenamefont {Luk}, \citenamefont {Decker}, \citenamefont {Neshev},
  \citenamefont {Brener},\ and\ \citenamefont {Kivshar}}]{Chong15}%
  \BibitemOpen
  \bibfield  {author} {\bibinfo {author} {\bibfnamefont {Katie~E.}\
  \bibnamefont {Chong}}, \bibinfo {author} {\bibfnamefont {Isabelle}\
  \bibnamefont {Staude}}, \bibinfo {author} {\bibfnamefont {Anthony}\
  \bibnamefont {James}}, \bibinfo {author} {\bibfnamefont {Jason}\ \bibnamefont
  {Dominguez}}, \bibinfo {author} {\bibfnamefont {Sheng}\ \bibnamefont {Liu}},
  \bibinfo {author} {\bibfnamefont {Salvatore}\ \bibnamefont {Campione}},
  \bibinfo {author} {\bibfnamefont {Ganapathi~S.}\ \bibnamefont {Subramania}},
  \bibinfo {author} {\bibfnamefont {Ting~S.}\ \bibnamefont {Luk}}, \bibinfo
  {author} {\bibfnamefont {Manuel}\ \bibnamefont {Decker}}, \bibinfo {author}
  {\bibfnamefont {Dragomir~N.}\ \bibnamefont {Neshev}}, \bibinfo {author}
  {\bibfnamefont {Igal}\ \bibnamefont {Brener}}, \ and\ \bibinfo {author}
  {\bibfnamefont {Yuri~S.}\ \bibnamefont {Kivshar}},\ }\bibfield  {title}
  {\enquote {\bibinfo {title} {Polarization-independent silicon metadevices for
  efficient optical wavefront control},}\ }\href {\doibase
  10.1021/acs.nanolett.5b01752} {\bibfield  {journal} {\bibinfo  {journal}
  {Nano Letters}\ }\textbf {\bibinfo {volume} {15}},\ \bibinfo {pages}
  {5369--5374} (\bibinfo {year} {2015})}\BibitemShut {NoStop}%
\bibitem [{\citenamefont {Shalaev}\ \emph {et~al.}(2015)\citenamefont
  {Shalaev}, \citenamefont {Sun}, \citenamefont {Tsukernik}, \citenamefont
  {Pandey}, \citenamefont {Nikolskiy},\ and\ \citenamefont
  {Litchinitser}}]{Shalaev15}%
  \BibitemOpen
  \bibfield  {author} {\bibinfo {author} {\bibfnamefont {Mikhail~I.}\
  \bibnamefont {Shalaev}}, \bibinfo {author} {\bibfnamefont {Jingbo}\
  \bibnamefont {Sun}}, \bibinfo {author} {\bibfnamefont {Alexander}\
  \bibnamefont {Tsukernik}}, \bibinfo {author} {\bibfnamefont {Apra}\
  \bibnamefont {Pandey}}, \bibinfo {author} {\bibfnamefont {Kirill}\
  \bibnamefont {Nikolskiy}}, \ and\ \bibinfo {author} {\bibfnamefont
  {Natalia~M.}\ \bibnamefont {Litchinitser}},\ }\bibfield  {title} {\enquote
  {\bibinfo {title} {High-efficiency all-dielectric metasurfaces for
  ultracompact beam manipulation in transmission mode},}\ }\href {\doibase
  10.1021/acs.nanolett.5b02926} {\bibfield  {journal} {\bibinfo  {journal}
  {Nano Letters}\ }\textbf {\bibinfo {volume} {15}},\ \bibinfo {pages}
  {6261--6266} (\bibinfo {year} {2015})}\BibitemShut {NoStop}%
\bibitem [{\citenamefont {Kruk}\ \emph {et~al.}(2016)\citenamefont {Kruk},
  \citenamefont {Wong}, \citenamefont {Pshenay-Severin}, \citenamefont
  {O'Brien}, \citenamefont {Neshev}, \citenamefont {Kivshar},\ and\
  \citenamefont {Zhang}}]{Kruk16}%
  \BibitemOpen
  \bibfield  {author} {\bibinfo {author} {\bibfnamefont {Sergey~S.}\
  \bibnamefont {Kruk}}, \bibinfo {author} {\bibfnamefont {Zi~Jing}\
  \bibnamefont {Wong}}, \bibinfo {author} {\bibfnamefont {Ekaterina}\
  \bibnamefont {Pshenay-Severin}}, \bibinfo {author} {\bibfnamefont {Kevin}\
  \bibnamefont {O'Brien}}, \bibinfo {author} {\bibfnamefont {Dragomir~N.}\
  \bibnamefont {Neshev}}, \bibinfo {author} {\bibfnamefont {Yuri~S.}\
  \bibnamefont {Kivshar}}, \ and\ \bibinfo {author} {\bibfnamefont {Xiang}\
  \bibnamefont {Zhang}},\ }\bibfield  {title} {\enquote {\bibinfo {title}
  {Magnetic hyperbolic optical metamaterials},}\ }\href {\doibase
  10.1038/ncomms11329} {\bibfield  {journal} {\bibinfo  {journal} {Nature
  Communications}\ }\textbf {\bibinfo {volume} {7}},\ \bibinfo {pages} {11329}
  (\bibinfo {year} {2016})}\BibitemShut {NoStop}%
\bibitem [{\citenamefont {Paniagua-Dom{\'\i}nguez}\ \emph
  {et~al.}(2016)\citenamefont {Paniagua-Dom{\'\i}nguez}, \citenamefont {Yu},
  \citenamefont {Miroshnichenko}, \citenamefont {Krivitsky}, \citenamefont
  {Fu}, \citenamefont {Valuckas}, \citenamefont {Gonzaga}, \citenamefont {Toh},
  \citenamefont {Kay}, \citenamefont {Luk'yanchuk},\ and\ \citenamefont
  {Kuznetsov}}]{Lukyanchuk16}%
  \BibitemOpen
  \bibfield  {author} {\bibinfo {author} {\bibfnamefont {Ram{\'o}n}\
  \bibnamefont {Paniagua-Dom{\'\i}nguez}}, \bibinfo {author} {\bibfnamefont
  {Ye~Feng}\ \bibnamefont {Yu}}, \bibinfo {author} {\bibfnamefont {Andrey~E.}\
  \bibnamefont {Miroshnichenko}}, \bibinfo {author} {\bibfnamefont {Leonid~A.}\
  \bibnamefont {Krivitsky}}, \bibinfo {author} {\bibfnamefont {Yuan~Hsing}\
  \bibnamefont {Fu}}, \bibinfo {author} {\bibfnamefont {Vytautas}\ \bibnamefont
  {Valuckas}}, \bibinfo {author} {\bibfnamefont {Leonard}\ \bibnamefont
  {Gonzaga}}, \bibinfo {author} {\bibfnamefont {Yeow~Teck}\ \bibnamefont
  {Toh}}, \bibinfo {author} {\bibfnamefont {Anthony Yew~Seng}\ \bibnamefont
  {Kay}}, \bibinfo {author} {\bibfnamefont {Boris}\ \bibnamefont
  {Luk'yanchuk}}, \ and\ \bibinfo {author} {\bibfnamefont {Arseniy~I.}\
  \bibnamefont {Kuznetsov}},\ }\bibfield  {title} {\enquote {\bibinfo {title}
  {Generalized brewster effect in dielectric metasurfaces},}\ }\href {\doibase
  10.1038/ncomms10362} {\bibfield  {journal} {\bibinfo  {journal} {Nature
  Communications}\ }\textbf {\bibinfo {volume} {7}},\ \bibinfo {pages} {10362}
  (\bibinfo {year} {2016})}\BibitemShut {NoStop}%
\bibitem [{\citenamefont {Gerbier}\ \emph {et~al.}(2006)\citenamefont
  {Gerbier}, \citenamefont {Widera}, \citenamefont {F\"olling}, \citenamefont
  {Mandel},\ and\ \citenamefont {Bloch}}]{gerbier_pra_2006}%
  \BibitemOpen
  \bibfield  {author} {\bibinfo {author} {\bibfnamefont {Fabrice}\ \bibnamefont
  {Gerbier}}, \bibinfo {author} {\bibfnamefont {Artur}\ \bibnamefont {Widera}},
  \bibinfo {author} {\bibfnamefont {Simon}\ \bibnamefont {F\"olling}}, \bibinfo
  {author} {\bibfnamefont {Olaf}\ \bibnamefont {Mandel}}, \ and\ \bibinfo
  {author} {\bibfnamefont {Immanuel}\ \bibnamefont {Bloch}},\ }\bibfield
  {title} {\enquote {\bibinfo {title} {Resonant control of spin dynamics in
  ultracold quantum gases by microwave dressing},}\ }\href {\doibase
  10.1103/PhysRevA.73.041602} {\bibfield  {journal} {\bibinfo  {journal} {Phys.
  Rev. A}\ }\textbf {\bibinfo {volume} {73}},\ \bibinfo {pages} {041602(R)}
  (\bibinfo {year} {2006})}\BibitemShut {NoStop}%
\bibitem [{SOM()}]{SOM}%
  \BibitemOpen
  \href@noop {} {}\bibinfo {note} {See Supplemental Material for technical
  details, which includes
  Refs.~\cite{Jenkins2012a,Lee16,SVI10,Jenkins_long16,Facchinetti16,Olmos13,mandel03,FleischhauerEtAlRMP2005,Jackson,Chomaz12,Javanainen17,Chomaz12,Facchinetti18,Javanainen19,Allen03}}\BibitemShut
  {NoStop}%
\bibitem [{\citenamefont {Svidzinsky}\ \emph {et~al.}(2010)\citenamefont
  {Svidzinsky}, \citenamefont {Chang},\ and\ \citenamefont {Scully}}]{SVI10}%
  \BibitemOpen
  \bibfield  {author} {\bibinfo {author} {\bibfnamefont {Anatoly~A.}\
  \bibnamefont {Svidzinsky}}, \bibinfo {author} {\bibfnamefont {Jun-Tao}\
  \bibnamefont {Chang}}, \ and\ \bibinfo {author} {\bibfnamefont {Marlan~O.}\
  \bibnamefont {Scully}},\ }\bibfield  {title} {\enquote {\bibinfo {title}
  {Cooperative spontaneous emission of $n$ atoms: Many-body eigenstates, the
  effect of virtual {L}amb shift processes, and analogy with radiation of $n$
  classical oscillators},}\ }\href {\doibase 10.1103/PhysRevA.81.053821}
  {\bibfield  {journal} {\bibinfo  {journal} {Phys. Rev. A}\ }\textbf {\bibinfo
  {volume} {81}},\ \bibinfo {pages} {053821} (\bibinfo {year}
  {2010})}\BibitemShut {NoStop}%
\bibitem [{\citenamefont {Needham}\ \emph {et~al.}(2019)\citenamefont
  {Needham}, \citenamefont {Lesanovsky},\ and\ \citenamefont
  {Olmos}}]{Needham19}%
  \BibitemOpen
  \bibfield  {author} {\bibinfo {author} {\bibfnamefont {Jemma~A}\ \bibnamefont
  {Needham}}, \bibinfo {author} {\bibfnamefont {Igor}\ \bibnamefont
  {Lesanovsky}}, \ and\ \bibinfo {author} {\bibfnamefont {Beatriz}\
  \bibnamefont {Olmos}},\ }\bibfield  {title} {\enquote {\bibinfo {title}
  {Subradiance-protected excitation transport},}\ }\href {\doibase
  10.1088/1367-2630/ab31e8} {\bibfield  {journal} {\bibinfo  {journal} {New
  Journal of Physics}\ }\textbf {\bibinfo {volume} {21}},\ \bibinfo {pages}
  {073061} (\bibinfo {year} {2019})}\BibitemShut {NoStop}%
\bibitem [{\citenamefont {Morice}\ \emph {et~al.}(1995)\citenamefont {Morice},
  \citenamefont {Castin},\ and\ \citenamefont {Dalibard}}]{Morice1995a}%
  \BibitemOpen
  \bibfield  {author} {\bibinfo {author} {\bibfnamefont {Olivier}\ \bibnamefont
  {Morice}}, \bibinfo {author} {\bibfnamefont {Yvan}\ \bibnamefont {Castin}}, \
  and\ \bibinfo {author} {\bibfnamefont {Jean}\ \bibnamefont {Dalibard}},\
  }\bibfield  {title} {\enquote {\bibinfo {title} {Refractive index of a dilute
  bose gas},}\ }\href {\doibase 10.1103/PhysRevA.51.3896} {\bibfield  {journal}
  {\bibinfo  {journal} {Phys. Rev. A}\ }\textbf {\bibinfo {volume} {51}},\
  \bibinfo {pages} {3896--3901} (\bibinfo {year} {1995})}\BibitemShut {NoStop}%
\bibitem [{\citenamefont {Ruostekoski}\ and\ \citenamefont
  {Javanainen}(1997)}]{Ruostekoski1997a}%
  \BibitemOpen
  \bibfield  {author} {\bibinfo {author} {\bibfnamefont {Janne}\ \bibnamefont
  {Ruostekoski}}\ and\ \bibinfo {author} {\bibfnamefont {Juha}\ \bibnamefont
  {Javanainen}},\ }\bibfield  {title} {\enquote {\bibinfo {title} {Quantum
  field theory of cooperative atom response: Low light intensity},}\ }\href
  {\doibase 10.1103/PhysRevA.55.513} {\bibfield  {journal} {\bibinfo  {journal}
  {Phys. Rev. A}\ }\textbf {\bibinfo {volume} {55}},\ \bibinfo {pages}
  {513--526} (\bibinfo {year} {1997})}\BibitemShut {NoStop}%
\bibitem [{\citenamefont {Javanainen}\ \emph {et~al.}(1999)\citenamefont
  {Javanainen}, \citenamefont {Ruostekoski}, \citenamefont {Vestergaard},\ and\
  \citenamefont {Francis}}]{Javanainen1999a}%
  \BibitemOpen
  \bibfield  {author} {\bibinfo {author} {\bibfnamefont {Juha}\ \bibnamefont
  {Javanainen}}, \bibinfo {author} {\bibfnamefont {Janne}\ \bibnamefont
  {Ruostekoski}}, \bibinfo {author} {\bibfnamefont {Bjarne}\ \bibnamefont
  {Vestergaard}}, \ and\ \bibinfo {author} {\bibfnamefont {Matthew~R.}\
  \bibnamefont {Francis}},\ }\bibfield  {title} {\enquote {\bibinfo {title}
  {One-dimensional modeling of light propagation in dense and degenerate
  samples},}\ }\href {\doibase 10.1103/PhysRevA.59.649} {\bibfield  {journal}
  {\bibinfo  {journal} {Phys. Rev. A}\ }\textbf {\bibinfo {volume} {59}},\
  \bibinfo {pages} {649--666} (\bibinfo {year} {1999})}\BibitemShut {NoStop}%
\bibitem [{\citenamefont {Sokolov}\ \emph {et~al.}(2011)\citenamefont
  {Sokolov}, \citenamefont {Kupriyanov},\ and\ \citenamefont
  {Havey}}]{Sokolov2011}%
  \BibitemOpen
  \bibfield  {author} {\bibinfo {author} {\bibfnamefont {I.~M.}\ \bibnamefont
  {Sokolov}}, \bibinfo {author} {\bibfnamefont {D.~V.}\ \bibnamefont
  {Kupriyanov}}, \ and\ \bibinfo {author} {\bibfnamefont {M.~D.}\ \bibnamefont
  {Havey}},\ }\bibfield  {title} {\enquote {\bibinfo {title} {Microscopic
  theory of scattering of weak electromagnetic radiation by a dense ensemble of
  ultracold atoms},}\ }\href {\doibase 10.1134/S106377611101016X} {\bibfield
  {journal} {\bibinfo  {journal} {Journal of Experimental and Theoretical
  Physics}\ }\textbf {\bibinfo {volume} {112}},\ \bibinfo {pages} {246--260}
  (\bibinfo {year} {2011})}\BibitemShut {NoStop}%
\bibitem [{\citenamefont {Javanainen}\ \emph {et~al.}(2014)\citenamefont
  {Javanainen}, \citenamefont {Ruostekoski}, \citenamefont {Li},\ and\
  \citenamefont {Yoo}}]{Javanainen2014a}%
  \BibitemOpen
  \bibfield  {author} {\bibinfo {author} {\bibfnamefont {Juha}\ \bibnamefont
  {Javanainen}}, \bibinfo {author} {\bibfnamefont {Janne}\ \bibnamefont
  {Ruostekoski}}, \bibinfo {author} {\bibfnamefont {Yi}~\bibnamefont {Li}}, \
  and\ \bibinfo {author} {\bibfnamefont {Sung-Mi}\ \bibnamefont {Yoo}},\
  }\bibfield  {title} {\enquote {\bibinfo {title} {Shifts of a resonance line
  in a dense atomic sample},}\ }\href {\doibase 10.1103/PhysRevLett.112.113603}
  {\bibfield  {journal} {\bibinfo  {journal} {Phys. Rev. Lett.}\ }\textbf
  {\bibinfo {volume} {112}},\ \bibinfo {pages} {113603} (\bibinfo {year}
  {2014})}\BibitemShut {NoStop}%
\bibitem [{\citenamefont {Sutherland}\ and\ \citenamefont
  {Robicheaux}(2016{\natexlab{b}})}]{Sutherland16forward}%
  \BibitemOpen
  \bibfield  {author} {\bibinfo {author} {\bibfnamefont {R.~T.}\ \bibnamefont
  {Sutherland}}\ and\ \bibinfo {author} {\bibfnamefont {F.}~\bibnamefont
  {Robicheaux}},\ }\bibfield  {title} {\enquote {\bibinfo {title} {Coherent
  forward broadening in cold atom clouds},}\ }\href {\doibase
  10.1103/PhysRevA.93.023407} {\bibfield  {journal} {\bibinfo  {journal} {Phys.
  Rev. A}\ }\textbf {\bibinfo {volume} {93}},\ \bibinfo {pages} {023407}
  (\bibinfo {year} {2016}{\natexlab{b}})}\BibitemShut {NoStop}%
\bibitem [{Note1()}]{Note1}%
  \BibitemOpen
  \bibinfo {note} {The light and atomic field amplitudes here refer to the
  slowly varying positive frequency components, where the rapid oscillations
  $\protect \qopname \relax o{exp}(-i \omega t)$ at the laser frequency have
  been factored out.}\BibitemShut {Stop}%
\bibitem [{\citenamefont {Lee}\ \emph {et~al.}(2016)\citenamefont {Lee},
  \citenamefont {Jenkins},\ and\ \citenamefont {Ruostekoski}}]{Lee16}%
  \BibitemOpen
  \bibfield  {author} {\bibinfo {author} {\bibfnamefont {Mark~D.}\ \bibnamefont
  {Lee}}, \bibinfo {author} {\bibfnamefont {Stewart~D.}\ \bibnamefont
  {Jenkins}}, \ and\ \bibinfo {author} {\bibfnamefont {Janne}\ \bibnamefont
  {Ruostekoski}},\ }\bibfield  {title} {\enquote {\bibinfo {title} {Stochastic
  methods for light propagation and recurrent scattering in saturated and
  nonsaturated atomic ensembles},}\ }\href {\doibase
  10.1103/PhysRevA.93.063803} {\bibfield  {journal} {\bibinfo  {journal} {Phys.
  Rev. A}\ }\textbf {\bibinfo {volume} {93}},\ \bibinfo {pages} {063803}
  (\bibinfo {year} {2016})}\BibitemShut {NoStop}%
\bibitem [{\citenamefont {Jackson}(1999)}]{Jackson}%
  \BibitemOpen
  \bibfield  {author} {\bibinfo {author} {\bibfnamefont {John~David}\
  \bibnamefont {Jackson}},\ }\href@noop {} {\emph {\bibinfo {title} {Classical
  Electrodynamics}}},\ \bibinfo {edition} {3rd}\ ed.\ (\bibinfo  {publisher}
  {Wiley, New York},\ \bibinfo {year} {1999})\BibitemShut {NoStop}%
\bibitem [{\citenamefont {Jenkins}\ \emph
  {et~al.}(2016{\natexlab{b}})\citenamefont {Jenkins}, \citenamefont
  {Ruostekoski}, \citenamefont {Javanainen}, \citenamefont {Jennewein},
  \citenamefont {Bourgain}, \citenamefont {Pellegrino}, \citenamefont
  {Sortais},\ and\ \citenamefont {Browaeys}}]{Jenkins_long16}%
  \BibitemOpen
  \bibfield  {author} {\bibinfo {author} {\bibfnamefont {S.~D.}\ \bibnamefont
  {Jenkins}}, \bibinfo {author} {\bibfnamefont {J.}~\bibnamefont
  {Ruostekoski}}, \bibinfo {author} {\bibfnamefont {J.}~\bibnamefont
  {Javanainen}}, \bibinfo {author} {\bibfnamefont {S.}~\bibnamefont
  {Jennewein}}, \bibinfo {author} {\bibfnamefont {R.}~\bibnamefont {Bourgain}},
  \bibinfo {author} {\bibfnamefont {J.}~\bibnamefont {Pellegrino}}, \bibinfo
  {author} {\bibfnamefont {Y.~R.~P.}\ \bibnamefont {Sortais}}, \ and\ \bibinfo
  {author} {\bibfnamefont {A.}~\bibnamefont {Browaeys}},\ }\bibfield  {title}
  {\enquote {\bibinfo {title} {Collective resonance fluorescence in small and
  dense atom clouds: Comparison between theory and experiment},}\ }\href
  {\doibase 10.1103/PhysRevA.94.023842} {\bibfield  {journal} {\bibinfo
  {journal} {Phys. Rev. A}\ }\textbf {\bibinfo {volume} {94}},\ \bibinfo
  {pages} {023842} (\bibinfo {year} {2016}{\natexlab{b}})}\BibitemShut
  {NoStop}%
\bibitem [{\citenamefont {Fleischhauer}\ \emph {et~al.}(2005)\citenamefont
  {Fleischhauer}, \citenamefont {Imamoglu},\ and\ \citenamefont
  {Marangos}}]{FleischhauerEtAlRMP2005}%
  \BibitemOpen
  \bibfield  {author} {\bibinfo {author} {\bibfnamefont {Michael}\ \bibnamefont
  {Fleischhauer}}, \bibinfo {author} {\bibfnamefont {Atac}\ \bibnamefont
  {Imamoglu}}, \ and\ \bibinfo {author} {\bibfnamefont {Jonathan~P.}\
  \bibnamefont {Marangos}},\ }\bibfield  {title} {\enquote {\bibinfo {title}
  {Electromagnetically induced transparency: Optics in coherent media},}\
  }\href {\doibase 10.1103/RevModPhys.77.633} {\bibfield  {journal} {\bibinfo
  {journal} {Rev. Mod. Phys.}\ }\textbf {\bibinfo {volume} {77}},\ \bibinfo
  {pages} {633--673} (\bibinfo {year} {2005})}\BibitemShut {NoStop}%
\bibitem [{\citenamefont {{Schelkunoff}}(1936)}]{Schelkunoff36}%
  \BibitemOpen
  \bibfield  {author} {\bibinfo {author} {\bibfnamefont {S.~A.}\ \bibnamefont
  {{Schelkunoff}}},\ }\bibfield  {title} {\enquote {\bibinfo {title} {Some
  equivalence theorems of electromagnetics and their application to radiation
  problems},}\ }\href {\doibase 10.1002/j.1538-7305.1936.tb00720.x} {\bibfield
  {journal} {\bibinfo  {journal} {The Bell System Technical Journal}\ }\textbf
  {\bibinfo {volume} {15}},\ \bibinfo {pages} {92--112} (\bibinfo {year}
  {1936})}\BibitemShut {NoStop}%
\bibitem [{\citenamefont {Alaee}\ \emph {et~al.}(2020)\citenamefont {Alaee},
  \citenamefont {Gurlek}, \citenamefont {Albooyeh}, \citenamefont
  {Mart\'{\i}n-Cano},\ and\ \citenamefont {Sandoghdar}}]{Alaee20}%
  \BibitemOpen
  \bibfield  {author} {\bibinfo {author} {\bibfnamefont {Rasoul}\ \bibnamefont
  {Alaee}}, \bibinfo {author} {\bibfnamefont {Burak}\ \bibnamefont {Gurlek}},
  \bibinfo {author} {\bibfnamefont {Mohammad}\ \bibnamefont {Albooyeh}},
  \bibinfo {author} {\bibfnamefont {Diego}\ \bibnamefont {Mart\'{\i}n-Cano}}, \
  and\ \bibinfo {author} {\bibfnamefont {Vahid}\ \bibnamefont {Sandoghdar}},\
  }\bibfield  {title} {\enquote {\bibinfo {title} {Quantum metamaterials with
  magnetic response at optical frequencies},}\ }\href {\doibase
  10.1103/PhysRevLett.125.063601} {\bibfield  {journal} {\bibinfo  {journal}
  {Phys. Rev. Lett.}\ }\textbf {\bibinfo {volume} {125}},\ \bibinfo {pages}
  {063601} (\bibinfo {year} {2020})}\BibitemShut {NoStop}%
\bibitem [{\citenamefont {Olmos}\ \emph {et~al.}(2013)\citenamefont {Olmos},
  \citenamefont {Yu}, \citenamefont {Singh}, \citenamefont {Schreck},
  \citenamefont {Bongs},\ and\ \citenamefont {Lesanovsky}}]{Olmos13}%
  \BibitemOpen
  \bibfield  {author} {\bibinfo {author} {\bibfnamefont {B.}~\bibnamefont
  {Olmos}}, \bibinfo {author} {\bibfnamefont {D.}~\bibnamefont {Yu}}, \bibinfo
  {author} {\bibfnamefont {Y.}~\bibnamefont {Singh}}, \bibinfo {author}
  {\bibfnamefont {F.}~\bibnamefont {Schreck}}, \bibinfo {author} {\bibfnamefont
  {K.}~\bibnamefont {Bongs}}, \ and\ \bibinfo {author} {\bibfnamefont
  {I.}~\bibnamefont {Lesanovsky}},\ }\bibfield  {title} {\enquote {\bibinfo
  {title} {Long-range interacting many-body systems with alkaline-earth-metal
  atoms},}\ }\href {\doibase 10.1103/PhysRevLett.110.143602} {\bibfield
  {journal} {\bibinfo  {journal} {Phys. Rev. Lett.}\ }\textbf {\bibinfo
  {volume} {110}},\ \bibinfo {pages} {143602} (\bibinfo {year}
  {2013})}\BibitemShut {NoStop}%
\bibitem [{\citenamefont {Mandel}\ \emph {et~al.}(2003)\citenamefont {Mandel},
  \citenamefont {Greiner}, \citenamefont {Widera}, \citenamefont {Rom},
  \citenamefont {H{\"a}nsch},\ and\ \citenamefont {Bloch}}]{mandel03}%
  \BibitemOpen
  \bibfield  {author} {\bibinfo {author} {\bibfnamefont {Olaf}\ \bibnamefont
  {Mandel}}, \bibinfo {author} {\bibfnamefont {Markus}\ \bibnamefont
  {Greiner}}, \bibinfo {author} {\bibfnamefont {Artur}\ \bibnamefont {Widera}},
  \bibinfo {author} {\bibfnamefont {Tim}\ \bibnamefont {Rom}}, \bibinfo
  {author} {\bibfnamefont {Theodor~W.}\ \bibnamefont {H{\"a}nsch}}, \ and\
  \bibinfo {author} {\bibfnamefont {Immanuel}\ \bibnamefont {Bloch}},\
  }\bibfield  {title} {\enquote {\bibinfo {title} {Controlled collisions for
  multi-particle entanglement of optically trapped atoms},}\ }\href {\doibase
  10.1038/nature02008} {\bibfield  {journal} {\bibinfo  {journal} {Nature}\
  }\textbf {\bibinfo {volume} {425}},\ \bibinfo {pages} {937--940} (\bibinfo
  {year} {2003})}\BibitemShut {NoStop}%
\bibitem [{\citenamefont {Chomaz}\ \emph {et~al.}(2012)\citenamefont {Chomaz},
  \citenamefont {Corman}, \citenamefont {Yefsah}, \citenamefont {Desbuquois},\
  and\ \citenamefont {Dalibard}}]{Chomaz12}%
  \BibitemOpen
  \bibfield  {author} {\bibinfo {author} {\bibfnamefont {L}~\bibnamefont
  {Chomaz}}, \bibinfo {author} {\bibfnamefont {L}~\bibnamefont {Corman}},
  \bibinfo {author} {\bibfnamefont {T}~\bibnamefont {Yefsah}}, \bibinfo
  {author} {\bibfnamefont {R}~\bibnamefont {Desbuquois}}, \ and\ \bibinfo
  {author} {\bibfnamefont {J}~\bibnamefont {Dalibard}},\ }\bibfield  {title}
  {\enquote {\bibinfo {title} {Absorption imaging of a quasi-two-dimensional
  gas: a multiple scattering analysis},}\ }\href {\doibase
  10.1088/1367-2630/14/5/055001} {\bibfield  {journal} {\bibinfo  {journal}
  {New Journal of Physics}\ }\textbf {\bibinfo {volume} {14}},\ \bibinfo
  {pages} {055001} (\bibinfo {year} {2012})}\BibitemShut {NoStop}%
\bibitem [{\citenamefont {Javanainen}\ \emph {et~al.}(2017)\citenamefont
  {Javanainen}, \citenamefont {Ruostekoski}, \citenamefont {Li},\ and\
  \citenamefont {Yoo}}]{Javanainen17}%
  \BibitemOpen
  \bibfield  {author} {\bibinfo {author} {\bibfnamefont {Juha}\ \bibnamefont
  {Javanainen}}, \bibinfo {author} {\bibfnamefont {Janne}\ \bibnamefont
  {Ruostekoski}}, \bibinfo {author} {\bibfnamefont {Yi}~\bibnamefont {Li}}, \
  and\ \bibinfo {author} {\bibfnamefont {Sung-Mi}\ \bibnamefont {Yoo}},\
  }\bibfield  {title} {\enquote {\bibinfo {title} {Exact electrodynamics versus
  standard optics for a slab of cold dense gas},}\ }\href {\doibase
  10.1103/PhysRevA.96.033835} {\bibfield  {journal} {\bibinfo  {journal} {Phys.
  Rev. A}\ }\textbf {\bibinfo {volume} {96}},\ \bibinfo {pages} {033835}
  (\bibinfo {year} {2017})}\BibitemShut {NoStop}%
\bibitem [{\citenamefont {Allen}\ \emph {et~al.}(2003)\citenamefont {Allen},
  \citenamefont {Barnett},\ and\ \citenamefont {Padgett}}]{Allen03}%
  \BibitemOpen
  \bibfield  {author} {\bibinfo {author} {\bibfnamefont {L.}~\bibnamefont
  {Allen}}, \bibinfo {author} {\bibfnamefont {S.M.}\ \bibnamefont {Barnett}}, \
  and\ \bibinfo {author} {\bibfnamefont {M.J.}\ \bibnamefont {Padgett}},\
  }\href {https://books.google.co.uk/books?id=Vf32PZXJ2gMC} {\emph {\bibinfo
  {title} {Optical Angular Momentum}}},\ Optics \& Optoelectronics\ (\bibinfo
  {publisher} {Taylor \& Francis},\ \bibinfo {year} {2003})\BibitemShut
  {NoStop}%
\end{thebibliography}
\end{document}


\title{Supplemental Material to \\``Optical Magnetism and Huygens' Surfaces in Arrays of Atoms Induced by Cooperative Responses''}

\author{K.~E.~Ballantine}
\author{J.~Ruostekoski}

\affiliation{Department of Physics, Lancaster University, Lancaster, LA1 4YB, United Kingdom}

\date{\today}

\maketitle



\renewcommand{\thesection}{S\Roman{section}}
\renewcommand{\thesubsection}{S\Roman{section}.\Alph{subsection}}
\renewcommand{\theequation}{S\arabic{equation}}
\renewcommand{\thefigure}{S\arabic{figure}}
\renewcommand{\bibnumfmt}[1]{[S#1]}
\renewcommand{\citenumfont}[1]{S#1}

In this supplemental material we briefly recap the basic relations for the electrodynamics of light and atoms, and further illustrate  the role of collective excitations for the example of synthesizing optical magnetism. 
We provide some additional details of the spherical harmonics used to decompose the far-field radiation and the scattered light and excitations of the Huygens' surface. We show an additional demonstration of the Huygens' surface in the generation of a single orbital angular momentum state with $\hbar$ angular momentum per photon.  

\section{Electrodynamics of light and atoms}

\subsection{Equations of motion}

In the main section we characterize the optical response of both the unit-cell and the many-atom array by writing the equations of motion as $\dot{\vec{b}} = i\mathcal{H}\vec{b}+\vec{F}$ where $\vec{b}$ is the vector of polarization amplitudes $\mathcal{P}_\sigma^{(j)}$ and $\vec{F}$ represents the external driving by the incident light. To see the origin of these equations we note that in the limit of low light intensity, the polarization amplitudes obey~\cite{Jenkins2012a,Lee16}
\begin{equation}\label{eq:Peoms}
\frac{d}{dt} \Pc_{\mu}^{(j)}  
  =  \left( i \Delta_\mu^{(j)} - \gamma \right)
  \Pc^{(j)}_\mu + i\frac{\xi}{\mathcal{D}}\hat{\vec{e}}_\mu^{\ast}\cdot\epsilon_0\vec{E}_{\rm ext}(\vec{r}_j),
\end{equation}
where $\xi=6\pi\gamma/k^3$, the single atom linewidth is $\gamma=\mathcal{D}^2k^3/6\pi\hbar\epsilon_0$, and $\Delta_\mu^{(j)}=\omega-\omega^{(j)}_\mu$ are the the detunings of the $m=\mu$ level of atom $j$ from resonance. 
The light and atomic field amplitudes here refer to the slowly varying positive frequency components, where the rapid oscillations $\exp(-i \omega t)$ at the laser frequency have been factored out.
Each amplitude $\mathcal{P}_\mu^{(j)}$ at position $\vec{r}_j$  is driven by the field,
\begin{equation}\label{eq:Eext}
\vec{E}_{\rm ext}(\vec{r}_j) = \cbE(\vec{r}_j) + \sum_{l\neq j} \vec{E}_s^{(l)}(\vec{r}_j),
\end{equation} 
which consists of the the incident field $\cbE(\vec{r})$ and the scattered field $
\epsilon_0 \vec{E}_s^{(l)}(\vec{r})=\mathsf{G}(\vec{r}-\vec{r}_l)\vec{d}_l
$ from the dipole moment $\vec{d}_l$ of each other atom $l$. The scattered field expression equals the usual positive-frequency component of the electric field from a
monochromatic dipole $\bd$,
given that the dipole resides at the origin and the field is observed
at $\br$ \cite{Jackson}:
\begin{align}
{\sf G}({\bf r})&\bd =
{1\over4\pi\epsilon_0}
\big\{ k^2(\hat{\bf n}\!\times\!\bd
)\!\times\!\hat{\bf n}{e^{ikr}\over r} \nonumber\\&+[3\hat{\bf n}(\hat{\bf
n}\cdot\bd)-\bd]
\big( {1\over r^3} - {ik\over r^2}\big) e^{ikr}
\big\}-{\bd\over3\epsilon_0}\,\delta({\bf r})\,,
\label{eq:DOL}
\end{align}
with $\hat{\bf n} = {{\bf r}/ r}$ and $k = {\omega / c}$.

Inserting Eq.~(\ref{eq:Eext}) into Eq.~(\ref{eq:Peoms}), with the dipole moment expressed in terms of the polarization as $\vec{d}_j=\mathcal{D}\sum_{\mu}\hat{\vec{e}}_\mu\mathcal{P}_\mu^{(j)}$, gives 
\begin{eqnarray}
\frac{d}{dt} \Pc_{\mu}^{(j)}  
  =  \left( i \Delta_\mu^{(j)} - \gamma \right)
  \Pc^{(j)}_\mu + i\xi\sum_{l\neq j}\mathcal{G}^{(jl)}_{\mu\nu}\Pc^{(l)}_\nu \nonumber \\ \label{eq:poleoms}
+  i\frac{\xi}{{\cal D}} \hat{\vec{e}}_{\mu}^{\ast} \cdot
  \epsilon_0 \cbE(\vec{r}_j),
\end{eqnarray} 
with $\mathcal{G}^{(jl)}_{\mu\nu}=\hat{\vec{e}}^*_\mu\cdot \mathsf{G}(\vec{r}_j-\vec{r}_l)\hat{\vec{e}}_\nu$. The linear equations of motion can then be written in matrix form as above.

This equation also describes the decay of a single photon excitation. The full quantum dynamics of the atomic system for a given initial excitation and in the absence of a driving laser follows from the quantum master equation for the many-atom density matrix $\rho$,
 \begin{equation}
\begin{multlined}
\label{eq:rhoeom}
\dot{\rho} = i\sum_{j,\nu}\Delta_{\nu}\left[\hat{\sigma}_{j\nu}^{+}\hat{\sigma}_{j\nu}^{-},\rho\right] +i\sum_{jl\nu\mu (l\neq j)}\Omega^{(jl)}_{\nu\mu}\left[\hat{\sigma}_{j\nu}^{+}\hat{\sigma}_{l\mu}^{-},\rho\right] \\
+\sum_{jl\nu\mu}\gamma^{(jl)}_{\nu\mu}\left(
2\hat{\sigma}^{-}_{l\mu}\rho\hat{\sigma}^{+}_{j\nu}-\hat{\sigma}_{j\nu}^{+}\hat{\sigma}_{l\mu}^{-}\rho -\rho\hat{\sigma}_{j\nu}^{+}\hat{\sigma}_{l\mu}^{-}\right) \,,
\end{multlined}
\end{equation}
where $\hat{\sigma}_{j\nu}^{+}=(\hat{\sigma}_{j\nu}^{-})^\dagger=\ket{e_{j\nu}}\bra{g_j}$ is the raising operator to the excited state $\nu$ on atom $j$.
The diagonal terms of the dissipative matrix are $\gamma^{(jj)}_{\nu\nu}=\gamma$, while the off-diagonal elements of the dissipation and interaction terms are given by the real and imaginary parts of
\begin{equation}
\Omega^{(jl)}_{\nu\mu}+i\gamma^{(jl)}_{\nu\mu}=
\xi {\cal G}^{(jl)}_{\nu\mu} \, .
\end{equation}
Restricting to the subspace of at most a single excitation, and assuming a pure initial state, the density matrix splits into one excitation and zero excitation parts,
\begin{equation}
\rho = \ket{\psi}\bra{\psi} + p_g\ket{G}\bra{G},
\end{equation}
where $\ket{\psi}$ represents a single excitation and can be expanded over the atomic sites,
\begin{equation}
\ket{\psi}= \sum_{j,\nu} \Pc^{(j)}_{\nu}(t)\,\hat{\sigma}^{+}_{j\nu}\ket{G},
\end{equation}
and $\ket{G}$ is the ground state.
Regarding single-particle expectation values, the dynamics can equally be described by the evolution of a vector $\vec{b}$ of these amplitudes $\Pc^{(j)}_{\nu}$~\cite{SVI10}, which obey the same equations as the classical polarization in the absence of drive, $\dot{\vec{b}}=i\mathcal{H}\vec{b}$.

\subsection{Collective excitation eigenmodes}

In both the case of the evolution of a single atomic excitation and the case of driven classical polarization, the optical response of the lattice is then characterized by the eigenvectors $\vec{v}_n$ and eigenvalues $\delta_n+i\upsilon_n$, of the evolution matrix $\mathcal{H}$~\cite{Jenkins_long16}. While the eigenmodes are not orthogonal, due to $\mathcal{H}$ being non-Hermitian, we find in our numerics that they always form a basis. Hence, the state at a time $t$ can be expanded in this basis $\vec{b}(t)=\sum_n c_n\vec{v}_n$. We define the quantity
\begin{equation}
L_j = \frac{|\vec{v}_j^T \vec{b}(t)|^2}{\sum_i|\vec{v}_i^T \vec{b}(t)|^2},
\end{equation}
as a measure of the relative occupation of each collective mode~\cite{Facchinetti16}.

\section{Eigenmodes of a single unit-cell}

\begin{figure}[htbp]
  \centering
   \includegraphics[width=\columnwidth]{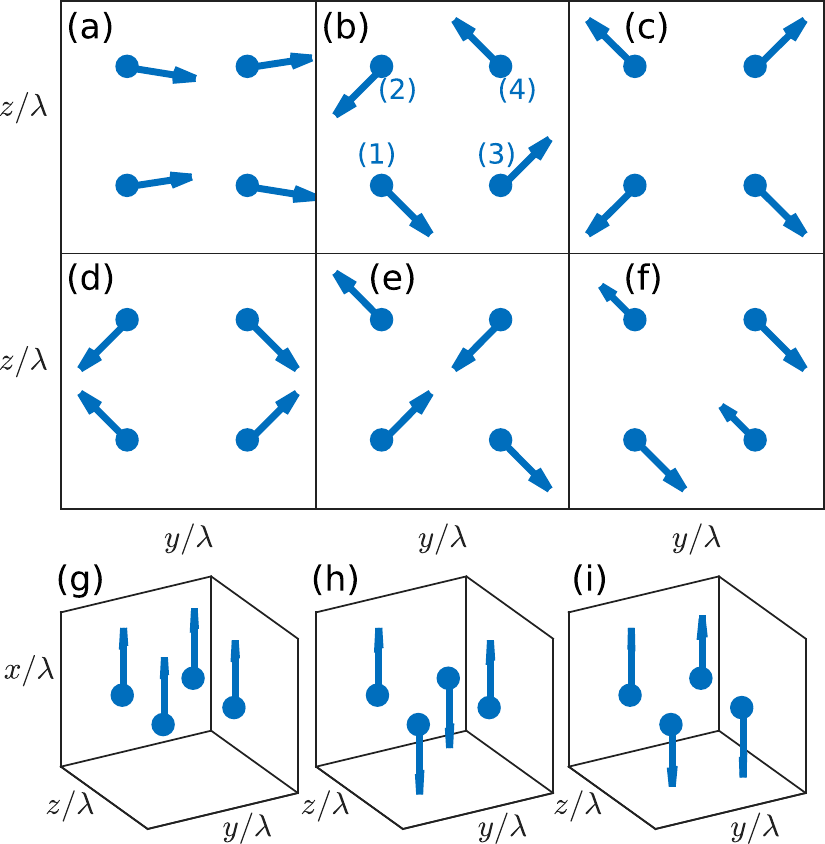}
  \caption{The orientation of the atomic dipoles for the eigenmodes of an isolated square unit-cell, with the same ordering as Table~1 of the main section, for the spacing $a=0.15\lambda$. Top and middle row: in-plane modes consisting of (a) electric dipole, (b) magnetic dipole ($n=4$), (c-e) electric quadrupole modes, and (f) mixed multipole character. Bottom row: modes with $x$ polarization consisting of (a) Electric dipole and (b-c) mixed multipole. Note the modes shown in (a), (f), and (i) are each doubly degenerate due to lattice symmetry. The ordering of the sites given in the text for the atomic level shifts is illustrated in (b).}
  \label{figmodes}
\end{figure}

We display in Fig.~\ref{figmodes} the excitation eigenmodes of a single isolated unit-cell of a planar array used to generate optically active magnetism [Fig.~1(a) in the main section]. Each unit-cell consists of four atoms and therefore
12 eigenmodes of which three are degenerate with some of the other modes and are obtained by trivial symmetry transformations of the dipole orientations. The properties of the eigenmodes are listed in Table 1 of the main section.

\section{Coupling to collective magnetic mode}

\begin{figure}[htbp]
  \centering
   \includegraphics[width=\columnwidth]{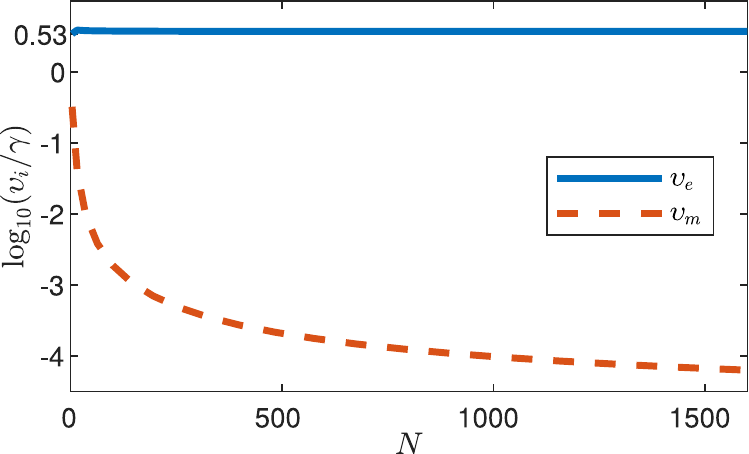}
  \caption{The collective linewidths $\upsilon_e$ and $\upsilon_m$ of the electric and magnetic dipole modes as a function of total atom number $N$ for a lattice with $a=0.15\lambda$ and $d=0.5\lambda$. While the linewidth of the magnetic mode depends strongly on $N$, the linewidth of the electric mode is almost constant, starting at $\upsilon_m=3.4$ for a single unit-cell and rising to $3.7$ at $N=1600$.
  }
  \label{fig:collectivelinewidth}
\end{figure}

The excitation of a collective mode consisting of magnetic dipoles in a planar array via the coupling between electric dipole eigenmodes (EDM) and magnetic dipole eigenmodes (MDM) is described by a two-mode model  [Eq.~(2) in the main section].
The incident field with the polarization $(\unitvec{e}_y+\unitvec{e}_z)/\sqrt{2}$ drives the EDM. The EDM in the $y$ and $z$ direction are clearly degenerate due to symmetry, and so the excited symmetric combination is also an eigenmode, with amplitude $\mathcal{P}_{e}$. When all the electronically excited levels are degenerate, this mode evolves independently of the MDM of amplitude $\mathcal{P}_{m}$ which is not excited.

When the level shifts are not equal ($\Delta_{\pm,0}^{(j)}\neq 0$), $\mathcal{P}_{e}$ and $\mathcal{P}_{m}$ no longer describe eigenmodes, and these are coupled together (as generally could the other eigenmodes of the $\Delta_{\pm,0}^{(j)} = 0$ system).  As shown in Fig.~\ref{figmodes}, the array of atoms in the $yz$ plane [Fig.~1(a) in the main section] has two separate families of modes, those with polarization in the $x$ direction out of the plane and those with polarizations in the plane. To avoid coupling the drive to those out of plane modes we choose $\Delta_{+}^{(j)}=\Delta_{-}^{(j)}$ on each atom. Then, to couple the EDM to the alternating phases of the $y$ and $z$ components of the MDM, we choose the remaining level shifts to have similarly alternating signs: 
\begin{subequations}
\begin{align}
 (\delta^{(\pm)}_{(4j+1)},\delta^{(\pm)}_{(4j+2)},\delta^{(\pm)}_{(4j+3)},\delta^{(\pm)}_{(4j+4)}) &= 
     \delta(2,0,2,0),\\
 (\delta^{(0)}_{(4j+1)},\delta^{(0)}_{(4j+2)},\delta^{(0)}_{(4j+3)},\delta^{(0)}_{(4j+4)}) &= 
     \delta(0,0,2,2),
\end{align}
\label{shifts}
\end{subequations}
where the ordering of the atoms is shown in Fig.~\ref{figmodes}(b).

The periodic variation of the level shifts can be produced by the ac Stark shift of an external standing-wave laser with the intensity varying along the principal axes of the array, such that the levels $m=\pm1$ are shifted using an intensity variation along the $z$ direction and the shift of the $m=0$ state with the intensity variation along the $y$ direction. In the both cases the intensity maxima are then separated by the distance $d$ between the adjacent unit-cells. Suitable transitions could be found, e.g., with Sr or Yb. For example, for the $^3P_0\rightarrow \mbox{}^3D_1$ transition of $^{88}$Sr the resonance wavelength $\lambda\simeq 2.6\mu$m and the linewidth $2.9\times 10^{5}$/s~\cite{Olmos13}. For the case of optical lattices the periodicity of the sinusoidal potential for the same transition with a magic wavelength may be chosen as 206.4nm and can also be modified to achieve the right periodicity by tilting the propagation direction of the lasers forming the lattice. Alternatively, atoms in different hyperfine states could occupy different lattice sites~\cite{mandel03}, with the associated description of the atom-light dynamics~\cite{Jenkins_long16,Lee16}, or the trapping potential strength for tweezers could possibly be spatially varied.

With this choice of level shifts given by Eq.~\eqref{shifts} we obtain [Eq.~(2) in the main section],
\begin{subequations}
\begin{align}
\label{zeeman1}
\partial_t \mathcal{P}_{e}^{(j)} &= (i\delta_e+i\Delta-\upsilon_e)\mathcal{P}_{e}^{(j)} + \delta\mathcal{P}_{m}^{(j)} +f, \\
\label{zeeman2}
\partial_t \mathcal{P}_{m}^{(j)} &= (i\delta_m+i\Delta-\upsilon_m)\mathcal{P}_{m}^{(j)} + \delta\mathcal{P}_{e}^{(j)}.
\end{align}
\label{zeemanboth}
\end{subequations}
The effective dynamics of Eqs.~\eqref{zeemanboth} can represent both a single unit-cell in isolation and the entire array of multiple unit-cells, but the collective resonance line shifts and linewidths of the EDM and MDM, $\delta_{e,m}$ and $\upsilon_{e,m}$, respectively, can considerably differ in the two cases, and generally vary with the number of unit-cells (Fig.~\ref{fig:collectivelinewidth}). The driving field is denoted by $f=i\xi\epsilon_0\e_0/\mathcal{D}$.  The alternating signs of $\Delta_{-,0}^{(j)}$ mean there is no coupling to other unit-cell eigenmodes with different symmetries. 

The steady state of Eqs.~\eqref{zeemanboth} is easily solved and we find the ratio of the MDM and EDM amplitudes to be
\begin{equation}
\left|\frac{\mathcal{P}_m}{\mathcal{P}_e}\right|=\left|\frac{\delta}{i\delta_m+i\Delta-\upsilon_m}\right|.
\end{equation}
On resonance, $\Delta=-\delta_m$, the ratio of amplitudes is determined by the collective linewidth $\upsilon_m$. As shown in Fig.~\ref{fig:collectivelinewidth} the linewidth of the MDM rapidly narrows as a function of the array size and becomes strongly subradiant. It is this collective resonance narrowing which allows the amplitude of the MDM to become large compared with the EDM amplitude, even for relatively small detuning strength $\delta$.

Equations~\eqref{zeemanboth} are similar to those describing the electromagnetically-induced transparency~\cite{FleischhauerEtAlRMP2005} of `dark' and `bright' states of noninteracting atoms in which case the atom population can be
trapped in the dark state. In the present case, the dark state is represented by a collective eigenmode, resulting from the resonant dipole-dipole interactions. It is this collective subradiant nature of the mode that drives the excitation into the MDM.

\section{Spherical harmonics}

 The multipole moments of the lattice unit-cells are characterized from the far-field radiation by decomposing the field in terms of vector spherical harmonics~\cite{Jackson} [Eq.~(1) in the main section].
The vector spherical harmonics are defined in terms of the ordinary spherical harmonics $Y_{lm}(\theta,\phi)$ as $\vec{\Psi}_{lm}=r\nabla Y_{lm}$ and $\vec{\Phi}_{lm}=\vec{r}\times \nabla Y_{lm}$ where $\vec{r}$ is the vector from the origin to the observation point, $\theta$ is the polar angle with the $x$ axis, and $\phi$ is the azimuthal angle from the $y$ axis in the $yz$ plane. They are orthogonal; $\int\vec{\Psi}_{lm}^\ast \vec{\Psi}_{l^\prime m^\prime}\mathrm{d}\Omega=\int\vec{\Phi}_{lm}^\ast \vec{\Phi}_{l^\prime m^\prime}\mathrm{d}\Omega=l(l+1)\delta_{l l^\prime}\delta_{m m^\prime}$, $\int\vec{\Psi}_{lm}^\ast \vec{\Phi}_{l^\prime m^\prime}\mathrm{d}\Omega=0$, and so the coefficients $\alpha$ can be found by projecting onto the corresponding vector harmonic.

\section{Huygens' surface}

For analyzing the properties of the Huygens' surface, the contributions of the both incident and scattered light are included. The scattered light can be calculated by summing up all the light scattered from all the atoms in the array.
We have verified that the transmission of light at distances $\lambda\alt \xi\ll \sqrt{{\cal A}}$ from a planar array of uniform excitations, where ${\cal A}$ denotes the area of the array, can also be estimated by~\cite{Chomaz12,Javanainen17,Facchinetti18,Javanainen19} 
\begin{align}
\epsilon_0 \vec{E} &= \e_0\hat{\vec{e}}_y e^{ik\xi} + \frac{ik}{2 {\cal A}}\sum_k \left[\vec{d}_k-\hat{\vec{e}}_x\cdot \vec{d}_k \hat{\vec{e}}_x\right]e^{i(\xi-x_k)},
\end{align}
where the second term denotes the scattered field $\vec{E}_S$ and the first term is the incident field. This expression has been used together with the microscopic calculation to analyze Huygens' surface, e.g., in Fig.~3 in the main section.

\begin{figure}[htbp]
  \centering
   \includegraphics[width=\columnwidth]{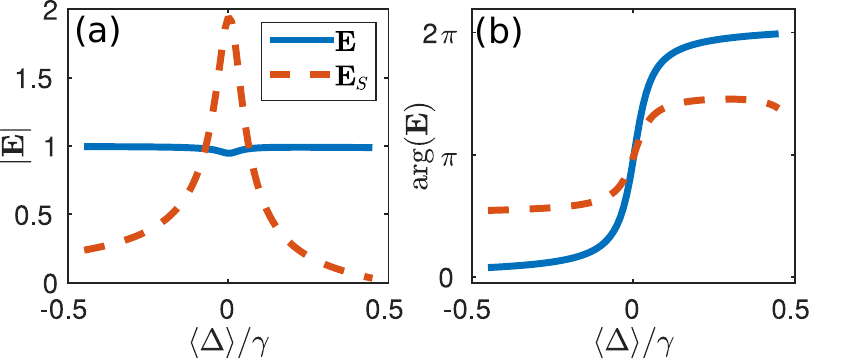}
  \caption{(a) The magnitude and (b) the phase of the total transmitted light and the scattered light for the same parameters, as Fig.~3(a) in the main section.
  }
  \label{fig:camparg}
\end{figure}

\begin{figure}[htbp]
  \centering
   \includegraphics[width=\columnwidth]{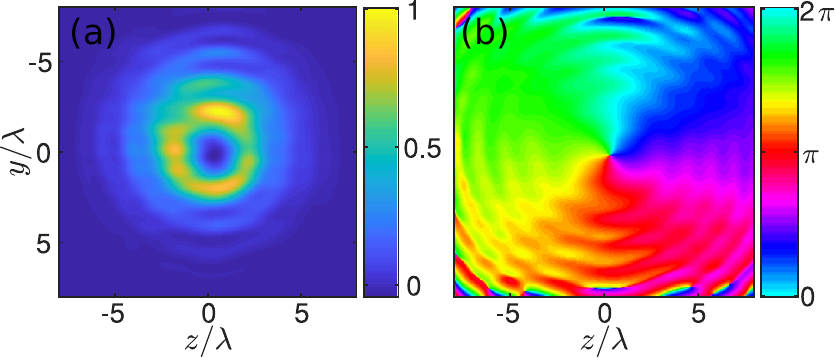}
  \caption{(a) The transmitted intensity and (b) phase $5\lambda$ from a $20\times 20$ Huygens' surface ($d=0.8\lambda$, $a=0.15\lambda$). An incident Gaussian beam with waist $5\lambda$ is transformed into an orbital angular momentum beam with angular $\hbar$ momentum per photon. Variations in the transmission are compensated by moving the center of the input beam a distance $\lambda$ in the $-y$ direction.
  }
  \label{fig:suppvortex}
\end{figure}

In Fig.~\ref{fig:camparg} we show the magnitude and the phase of the total transmitted field, and also the contribution of the scattered field alone. Fig.~\ref{fig:camparg}(b) shows that while the phase of the scattered field covers a range of $\pi$, the total field has a full range of $2\pi$. The contribution from electric and magnetic dipoles in a Huygens' surface add to give a scattered field with magnitude up to twice the incident field, as shown in Fig.~\ref{fig:camparg}(a), allowing for close to total transmission even when the scattered field is $\pi$ out of phase with the incident light.

In the main text we demonstrate the Huygens' surface by transforming a Gaussian into a superposition of orbital angular momentum (OAM) states. The surface can also be used to create single OAM states, where $\vec{E}\propto\exp{(il\phi)}$ where $\phi$ is the azimuthal angle in the plane, with OAM $l\hbar$ per photon~\cite{Allen03}. The resulting intensity and phase is shown in Fig.~\ref{fig:suppvortex} for $l=1$, with a characteristic $2\pi$ phase winding around the center.